\documentclass[a4paper,USenglish,cleveref, autoref, thm-restate,runningheads]{llncs}

\newcommand{\longversion}[1]{#1}
\newcommand{\shortversion}[1]{}
\usepackage[dvipsnames]{xcolor}
\usepackage{todonotes}
\usepackage{tikz}
\usetikzlibrary{arrows,shapes,snakes,automata,backgrounds,petri}
\usetikzlibrary{shapes.multipart}
\usepackage{comment}
\usepackage[noend]{algpseudocode}
\usepackage{caption}
\usepackage{subcaption}

\usepackage[utf8]{inputenc}          
\usepackage[T1]{fontenc}             
\usepackage[USenglish]{babel}        
\usepackage[intlimits]{amsmath}      
\usepackage{amsfonts}                
\usepackage{amssymb}                 
\usepackage{amsmath}                 

\usepackage{mathtools}
\usepackage{setspace}
\usepackage{graphicx}
\usepackage{hyperref}
\usepackage{dsfont}
\usepackage{xspace}

\usepackage{algorithm}
\usepackage[noend]{algpseudocode}

\usepackage{algorithm}
\usepackage[noend]{algpseudocode}

\newcommand{\EPRD}{\textsc{Extension Perfect Roman Domination}\xspace}

\newcommand{\URRD}{\textsc{Unique Response Roman Domination}\xspace}
\newcommand{\PRD}{\textsc{Perfect Roman Domination}}

\newcommand{\NP}{\textsf{NP}}
\newcommand{\FPT}{\textsf{FPT}}
\newcommand{\XP}{\textsf{XP}\xspace}
\newcommand{\W}[1]{\ensuremath{\textsf{W}[#1]}}
\newcommand{\paraNP}{\textsf{para-NP}\xspace}
\newcommand{\Active}{A}

\newcommand{\rdf}{\longversion{Roman dominating function}\shortversion{Rdf}\xspace}
\newcommand{\rdfs}{\longversion{Roman dominating functions}\shortversion{Rdfs}\xspace}

\newcommand{\prdf}{\longversion{perfect Roman dominating function}\shortversion{pRdf}\xspace}
\newcommand{\prdfs}{\longversion{perfect Roman dominating functions}\shortversion{pRdfs}\xspace}

\newcommand{\urrdf}{\longversion{\longversion{unique response Roman dominating function}\shortversion{urRdf}}\shortversion{urRdf}\xspace}
\newcommand{\urrdfs}{\longversion{\longversion{unique response Roman dominating functions}}\shortversion{urRdfs}\xspace}

\newcommand{\mRdfSet}[1]{\mu-\mathcal{RDF} \left(#1\right)}
\newcommand{\mpRdfSet}[1]{\mu-\mathcal{PRDF} \left(#1\right)}

\newcommand{\no}{\textsf{no}\xspace}
\newcommand{\yes}{\textsf{yes}\xspace}

\newcommand{\RomanUpperbound}{1.9332}
\newcommand{\RomanLowerbound}{1.7441}

\newcommand{\Oh}{\mathcal{O}}
\newcommand{\weightfun}{\omega}
\newcommand{\weightfunction}[1]{\weightfun\left(#1\right)}

\newcommand{\iffl}{if\longversion{ and only i}f\xspace}
\newtheorem{observation}{Observation}

\newtheorem{brarule}{Branching Rule}

\newtheorem{redrule}{Reduction Rule}

\newenvironment{pf}{\begin{proof}}{\hfill\qed\end{proof}}\newenvironment{pfclaim}{
\emph{Proof.}}
{\hfill$\Diamond$}

\begin{document}
\title{Perfect Roman Domination and Unique Response Roman Domination}
\author{Henning Fernau
\orcidID{0000-0002-4444-3220}
\and Kevin Mann
\orcidID{0000-0002-0880-2513} 
}
\authorrunning{Kevin Mann et al.}
\institute{
Universit\"at Trier, Fachbereich~4 -- Abteilung Informatikwissenschaften\\  
54286 Trier, Germany.\\
\email{\{fernau,mann\}@uni-trier.de}
}
\maketitle

\begin{abstract}
 The idea  of enumeration algorithms with polynomial delay is to polynomially bound the running time between any two subsequent solutions output by the enumeration algorithm.  While it is open for more than four decades if all minimal dominating sets of a graph can be enumerated in output-polynomial time, it has recently been proven that pointwise-minimal Roman dominating functions can be enumerated even with polynomial delay. The idea of the enumeration algorithm was to use polynomial-time solvable extension problems. 

 We use this as a motivation to prove that also two variants of Roman dominating functions studied in the literature, named perfect and unique response, can be enumerated with polynomial delay. This is interesting since \textsc{Extension Perfect Roman Domination} is \W{1}-complete if parameterized by the weight of the given function and even \W{2}-complete if parameterized by the number vertices assigned 0 in the pre-solution, as we prove. Otherwise, efficient solvability of extension problems and enumerability with polynomial delay tend to go hand-in-hand.
 We achieve our enumeration result by constructing a bijection to Roman dominating functions, where the corresponding extension problem is polynomimal-time solvable. 
 Furthermore, we show that \textsc{Unique Response Roman Domination} is solvable in polynomial time on split graphs, while \textsc{Perfect Roman Domination} is \NP-complete on this graph class, which proves that both variations, albeit coming with a very similar definition, do differ in some complexity aspects. This way, we also solve an open problem from the literature.
\end{abstract}
    
\section{Introduction}

\subsection{Roman Domination} 
Historically, \textsc{Roman Domination} is motivated by the defense strategy of the Roman Empire. The idea was to position the armies on regions in such a way that either (1) there is one army in this region or (2) there are two armies on one neighbored region. Translated to graphs we map each vertex to 0, 1 or 2. Such a function is called a Roman dominating function if each vertex with value 0 has a neighbor of value 2.  \textsc{Roman Domination} has as input a graph~$G$ and a positive number~$k$, and the question is if there exists a Roman dominating function such that the sum of the values of all vertices is at most~$k$. In the last decades, this problem received notable attention \cite{Ben2004,ChaCCKLP2013,Dre2000a,Fer08,Lie2007,Lieetal2008,LiuCha2013,Pagetal2002,PenTsa2007,ShaWanHu2010}.

As for dominating set, there are also many variants of Roman dominating functions which were considered in the literature. Examples are Roman-$\{2\}$-domination (also know as Italian domination)~\cite{CheHHM2016}, double Roman domination \cite{AbdCheShe2017,BanHenPra2020,BeeHayHed2016} and 
total Roman domination~\cite{AbdHSY2016}.

\subsection{Perfect Variations}
Here, we will consider two further variations of Roman domination, perfect and unique response Roman domination. A \emph{perfect Roman dominating function} (introduced by Henning \emph{et al.} \cite{HenKloMac2018}) is a Roman dominating function where each vertex with value 0 has \emph{exactly} one neighbor with value~2. If such a function additionally satisfies that all vertices with at least~1 as value have no neighbor with value 2, then it is a \emph{unique response Roman dominating function} (introduced by Rubalcaba and Slater \cite{RubSla2007}). Both variations can be seen as a way to translate the idea of perfect domination into the realm of Roman dominating functions. As a further motivation, we can also consider the idea of positioning armies on regions: If the armies are placed according to a perfect Roman dominating function and a region without any army is attacked, then it is clear from which region an army moves to secure the attacked region, so no time is wasted to first agree on who is the one to take action and move to the endangered region.

Supplementing results from the literature, we study the underlying minimization problems on split and on cobipartite graphs, which shows that these two seemingly very similar notions give raise to a different complexity behavior. However, the main focus of this paper is on enumeration, both from an input-sensitive and from an output-sensitive perspective.

\subsection{Enumeration}
Enumeration is wide area of research, as also testified by specialized workshops like~\cite{FerGolSag2018}. For some examples, we refer to the survey~\cite{Was2016}. From a practical point of view, enumeration can be interesting if not all aspects of the problem have been satisfyingly modeled. For instance, it is possible to enumerate all (inclusion-wise) minimal dominating sets of a graph of order~$n$ in time  $\Oh(1.7159^n)$ \cite{Fometal2008a}.  As here only the size of the input graph is taken into consideration, one also speaks of an input-sensitive analysis. In contrast to classical complexity, one can often also give lower-bound examples, which are families of graphs that possess, when taking an $n$-vertex representative thereof, $\Omega(1.5704^n)$ many minimal dominating sets \cite{Fometal2008a}. When lower and upper bounds match, we can consider this type of enumeration as being optimal. However, even such an optimal enumeration algorithm can be dissatisfying for a user, as she might have to wait exponential time between two output solutions.  This motivates to study enumeration algorithms from an output-sensitive perspective. The most important notions have been introduced by D.~S. Johnson \emph{et al.}~\cite{JohPapYan88a}. We focus on the most restricted variant. An enumeration algorithm has \emph{polynomial delay} if the time between two subsequent outputs of the algorithm can be bounded by some polynomial, as well as the time from the start to the first output and the time between the last output and the termination of the algorithm. 
This is a very desirable  property if two processes work as in a production line: one process generates the solutions, while the other one works on the generated solution. After finishing its work on one solution, the second process does not want to work for a long time to start working on the next solution.
It is open since decades if all minimal dominating sets can be enumerated with polynomial delay.

We motivate the enumeration of perfect/unique response Roman dominating functions by the main result of F. N. Abu-Khzam \emph{et al.}~\cite{AbuFerMan2022}. There, it is proven that all minimal Roman dominating functions of a graph of order~$n$ can be enumerated in time $\mathcal{O}(\RomanUpperbound^n)$ with polynomial delay. In this case, minimality is defined with respect to a pointwise order, \emph{i.e.}, for $f,g: V \to\{0,1,2\}$, $f\leq g$ \iffl $f(v)\leq g(v)$ for all $v\in V$. 
To ensure polynomial delay for enumerating minimal Roman dominating functions, F. N. Abu-Khzam \emph{et al.} used polynomial-time solvable extension problems.  

\subsection{Extension Problems}
For a general definition of extension problems, we refer to~\cite{CasFGMS2021}. Here, we will only discuss the extension version of minimization problems on graphs. Therefore, we can be more specific. Depending on the concrete problem (we choose to illustrate this in the following with the classical problem \textsc{Dominating Set} in parentheses), a graph $G=(V,E)$ defines the search space $\textsf{presol}(G)$ of \emph{pre-solutions} (in our example, $\textsf{presol}(G)=2^{V(G)}$) and a set of \emph{solutions} $\textsf{sol}(G)\subseteq \textsf{presol}(G)$ (dominating sets). For the extension version, we also need to define a partial order $\preceq$ on  $\textsf{presol}(G)$ (which is $\subseteq$ for domination). The notion of a \emph{minimal solution} is understood with respect to~$\preceq$: $s\in \textsf{sol}(G)$ is called minimal if, for each $p\in \textsf{presol}(G)\setminus\{s\}$, $p\preceq s$ implies $p\notin \textsf{sol}(G)$.  An instance of the extension version consists, apart from the graph~$G$, in a pre-solution $p\in \textsf{presol}(G)$ (some set of vertices in our example). The question is if there exists a minimal solution $s\in \textsf{sol}(G)$ with $p \preceq s$. Notice that a typical branching algorithm used for the enumeration of all minimal solutions will implicitly create a pre-solution~$p$, and then efficiently determining if any minimal solution exists that extends~$p$ would be very beneficial. 
In the special case of $\textsc{Extension Roman Domination}$, $\textsf{presol}(G)$ is the set of all mappings  $f: V \to \{0,1,2\}$ and $\textsf{sol}(G)$ is the set of all Roman dominating functions. Now, $\preceq$ is given by the partial order $\leq$ described above: $f\leq g$ holds if $f(v) \leq g(v)$ for each $v \in V$.  

\subsection{Organization of the Paper}
 We start by introducing important notations and definitions in \autoref{sec:pre}. Then we consider the optimization problems \textsc{Perfect Roman Domination} and \textsc{Unique Response Roman Domination} on split and cobipartite graphs. We show that \textsc{Perfect Roman Domination} is \NP-complete while \textsc{Unique Response Roman Domination} is polynomial-time solvable. To our knowledge, this is the first graph class where this is shown. Section \ref{sec:Enum_urrdf} provides a polynomial-delay enumeration algorithm for unique response Roman dominating functions.  In \autoref{sec:ExtPrd} we demonstrate that \textsc{Extension Perfect Roman Domination} is \NP-complete, \W{1}-complete when parameterized by the pre-solution size and \W{2}-complete when parameterized by the number of vertices assigned 0 by the pre-solution. Nonetheless, we present a way to enumerate all minimal perfect Roman dominating functions with polynomial delay in \autoref{sec:Enum_minPrd} by showing a one-to-one correspondence between minimal perfect Roman dominating functions and minimal Roman dominating functions. 

\section{Preliminaries}\label{sec:pre}

\subsection{General Notions}
Let $\mathbb{N}$ denote the set of all nonnegative integers (including 0). For $n\in \mathbb{N}$, we will use the notation $[n]\coloneqq\{1,\ldots,n\}$. Let $G=(V,E)$ be a graph. $N_G(v)$ describes the \emph{open neighborhood} of $v\in V$ with respect to $G$. The \emph{closed neighborhood} of $v\in V$  with respect to $G$ is defined by $N_G[v] \coloneqq N_G(v)\cup \{ v \}$. For a set $A\subseteq V$ the open neighborhood is defined as $N_G(A) \coloneqq \left(\bigcup_{v\in A} N_G(v)\right)$. The closed neighborhood of $A$ is given by $N_G[A]\coloneqq N_G(A) \cup A$. Furthermore, the private neighborhood of a vertex $v\in A$ with respect to $G$ and $A$ is denoted by $P_{G,A}(v)\coloneqq N_G[v] \setminus N_G[A\setminus \{v\}]$.

Let $G=(V,E)$ be a graph. 
We say $D\subseteq V$ is a dominating set if $N[D]=D$.
A set $D\subseteq V$ is called \emph{perfect dominating} if $D$ is a dominating set and for all $v,u \in D$ with $v \neq u$, $ N[v]\cap N[u]=\emptyset$.

Let $A,B$ be two sets. $B^A$ is the set of all function $f:A\to B$.
For a functions $f:A\rightarrow \mathbb{N}$ on a finite set $A$, we define $\weightfun(f)=\sum_{a\in A}f(a)$. For $A\subseteq B$ $\chi_A:B\to \mathbb{N}$ denotes the characteristic function ($\chi(a)=1$ \iffl $a\in A$; $\chi(a)=0$ otherwise) 

\subsection{Basic Decision Problems}

\centerline{\fbox{\begin{minipage}{.99\textwidth}
\textbf{Problem name: }\textsc{Unique Response Roman Domination}\\
\textbf{Given: } A graph $G=(V,E)$ and $k\in \mathbb{N}$\\
\textbf{Question: } Is there a \urrdf~$f$ on~$G$ with $\weightfun(f)\leq k$?\end{minipage}
}}

\centerline{\fbox{\begin{minipage}{.99\textwidth}
\textbf{Problem name: }\textsc{Perfect Roman Domination}\\
\textbf{Given: } A graph $G=(V,E)$ and $k\in \mathbb{N}$\\
\textbf{Question: } Is there a \prdf~$f$ on~$G$ with $\weightfun(f)\leq k$?\end{minipage}
}}

Let $u_R(G)$ denote the smallest weight of any \urrdf~$f$ on~$G$. Furthermore, $\gamma_R^p(G)$ denotes the smallest weight of any \prdf on $G$.
The following is known about the complexity of these decision problems.
\begin{itemize}
    \item \URRD\ is \NP-complete even on regular bipartite  graphs \cite{CabPueRod2021}. Furthermore, Banerjee \emph{et al.} \cite{BanChaPra2023} showed that this problem is \NP-complete on chordal graphs, and polynomial-time solvable on distance-hereditary and interval graphs. They also prove that there is no \URRD polynomial-time approximation algorithm within a factor of $n^{1-\epsilon}$ for any constant $\epsilon>0$ and any input graph of order $n$, unless $\NP=\textsf{P}$.
    \item \PRD\ is \NP-complete on chordal graphs, planar graphs, and bipartite graphs and polynomial time solveable on block graphs, cographs, series-parallel graphs, and proper interval graphs \cite{BanKeiPra2019}.
\end{itemize}

\subsection{Characterizing Perfect Roman Dominating Functions}
Consider a graph~$G=(V,E)$ and a function $f: V\rightarrow \{ 0,1,2\}$. We define $V_i(f):=\{v\in V\mid f(v)=i\}$ for each $i\in \{0,1,2\}$. If $v\in V$ obeys $f(v)=2$, then $u\in N_G(v)$ with $f(u)=0$ is called a \emph{private neighbor} of~$v$ (with respect to~$f$) if $\vert N_G(u)\cap V_2(f)\vert =1$.

\begin{observation}\label{obs:private-URRD}
Let $G=(V,E)$ be a graph and $f: V\rightarrow \{ 0,1,2\}$.
\begin{enumerate}
    \item If $f$ is a \prdf, then every neighbor $u$ with $f(u)=0$ of some $v$ with $f(v)=2$ is a private neighbor of~$v$.
    \item If $f$ is a Roman dominating function such that every neighbor $u$ with $f(u)=0$ of some $v$ with $f(v)=2$ is a private neighbor of~$v$, then $f$ is a \prdf.
\end{enumerate}

\end{observation}

\begin{pf}Let us first prove the first item. 
If it were not true, then there would be some neighbor $u$ with $f(u)=0$ of some $v$ with $f(v)=2$ that is non-private, \emph{i.e.}, there exists some $v'\in N_G(u)$, $v'\neq v$, with $f(v')=2$. This contradicts the assumption that  $f$ is a \prdf. 
To see the second implication, observe that if $f$ is a Roman domination function that is not perfect, then there must be a vertex $u$ with $f(u)=0$ such that $u$ has two neighbors $v_1,v_2$ with $f(v_1)=f(v_2)=2$. Hence, $u$ is not a private neighbor of~$v_1$.
\end{pf}

\section{Optimization on Split and Cobipartite Graphs}\label{sec:urrd_prd_on_split}

A split graph  $G=(V,E)$ is a graph such that the vertex set $V$ can be partitioned into $C,I$, where $C$ is a clique and $I$ is an independent set of~$G$. It seems to be open in which complexity class \textsc{Perfect Roman Domination} and  \textsc{Unique Response Roman Domination} are on split graphs, as explicitly asked in \cite{BanChaPra2023}. The main result of this section will give the answers.

\subsection{Unique Response Roman Domination on Split Graphs}
\begin{lemma}\label{lem:split-URRD}
Let $G=(V,E)$ be a connected split graph with a clique~$C$ and an independent set $I$ such that $V=C\cup I$ and $C \cap I=\emptyset$. For each \urrdf $f:V\rightarrow \{0,1,2\}$, one of the following conditions holds:
\begin{itemize}
    \item $V_2(f)\cap I=\emptyset$ and $\vert V_2(f)\cap C\vert \leq 1$; or
    \item $V_2(f) \subseteq I$.
\end{itemize}
\end{lemma}

\begin{pf} Suppose that $V_2(f)$ is not a subset of~$I$.
By the definition of \urrdf, two vertices of~$C$ cannot have the value~2, since $C$ is a clique. Hence, $\vert V_2(f)\cap C\vert \leq 1$. For the sake of contradiction, assume there are $v\in I$ and $u\in C$ with $f(u)= f(v) =2$. As $f$ is a \urrdf, $u$ and $v$ cannot be neighbors. Since $\emptyset\subsetneq N_G(v)\subseteq C$ holds as $G$ is connected, $v$ has a neighbor $w\in C$. This contradicts the assumption on $f$, as $\vert N_G(w)\cap V_2(f)\vert\geq 2$.
\end{pf}

Let us take a look at the \urrdf for the two cases of \autoref{lem:split-URRD}. Without loss of generality, we can assume that for the given split graph $G=(V,E)$, we have a decomposition $V=C\cup I$ into a clique~$C$ and an independent set~$I$ such that $C$ is inclusion-wise maximal. Consider the \urrdf $f:V\rightarrow \{0,1,2\}$.

First assume~$f$ fulfills $V_2(f)\cap I=\emptyset$ and $\vert V_2(f)\cap C\vert \leq 1$. Since $V_2(f)\cap C = \emptyset$ would imply $f=1$ (constant) and $\weightfun(f)=\vert C\vert + \vert I\vert$, let us assume $\{v\} = V_2(f)\cap C$. Therefore, $C\cap V_1(f)=\emptyset$ holds. Hence, $\weightfun(f)=2+\vert I\setminus N_G(v)\vert$. This implies $u_R(G)\leq \min_{v\in C} \left(2 + \vert I\setminus N_G(v)\vert\right)\leq 2 + \vert I\vert$. 
 
Secondly, let us assume $V_2(f) \subseteq I$. As $I$ is an independent set and there is no vertex in $C$ with the value 2, $f(v)\geq 1$ for each $v\in I$. This implies 
$$\weightfun(f)= \vert I\vert + \vert  V_2(f)\vert +\vert C\setminus N_G(V_2(f)) \vert  \geq \vert I\vert + 1.$$ Equality holds if and only if there exists a vertex $v\in I$ with $N_G(v)=C$ and, for all $u\in I$, $f(u)=2\iff u=v$. In this case $C$ is not maximal, contradicting our assumption. Thus, this second case can never give a smaller value than the first one, \emph{i.e.}, $u_R(G)\leq \min_{v\in C}\left( 2 + \vert I\setminus N_G(v)\vert\right)$. In order to solve \textsc{Unique Response Roman Domination} on~$G$, we only have to find the vertex with the highest degree, as it will be in a maximal clique.
 
 \begin{corollary}\label{cor:URRDF_P_on_split}
 \URRD can be solved in time $\Oh(n+m)$ on split graphs.
 \end{corollary}

\subsection{Perfect Roman Domination on Split Graphs}

Interestingly, the seemingly very similar problem \PRD on split graphs is harder to solve. To see this, we will use a problem called \textsc{Perfect Domination} that is defined next.

\centerline{\fbox{\begin{minipage}{.99\textwidth}
\textbf{Problem name: }\textsc{Perfect Domination}\\
\textbf{Given: } A graph $G=(V,E)$ and $k\in \mathbb{N}$\\
\textbf{Question: } Is there a perfect dominating set~$D$ on~$G$ with $\vert D \vert \leq k$?\end{minipage}
}}
It is known that \PRD is \NP-complete, see \cite{FelHoo91}.
 
 \begin{theorem}
 \textsc{Perfect Roman Domination} on split graphs is \NP-complete.
 \end{theorem}
\begin{pf}
Membership follows from  \PRD\ on general graphs.
To prove \NP-hardness, we will use \textsc{Perfect Domination}. Let $G=(V,E)$ be a graph and $k\in \mathbb{N}$. To avoid trivialities, $k\leq |V|$.

Define $t= 3 \cdot \vert V\vert + 4$, $k'=k + \vert V\vert + 4$ and $G'=(V',E')$ with 
\begin{equation*}
    \begin{split}
        V' =& \{x,y\}\cup\{v',v_i\mid v\in V, i\in [t]\} \cup\{a_1,b_1,\ldots,a_t,b_t\},\\
        E' =& \{\{x,a_i\},\{y,b_i\}\mid i\in \left[t\right]\}\cup \{\{v',u_i\} \mid v,u\in V, u \in N[v], i\in [t]\}\cup\\
        &\binom{\{x,y\}\cup \{v' \mid v\in V\}}{2} .
    \end{split}
\end{equation*}

The instance $(G',k')$ of \PRD\ can be constructed in polynomial time. Moreover, $G'$ is easily seen to be a split graph. We still have to prove that $(G,k)$ is a \yes-instance of \textsc{Perfect Domination} \iffl $(G',k')$ is a \yes-instance of \PRD.

Let $S \subseteq V$ be a perfect dominating set of~$G$ with $\vert S\vert \leq k$. Define $f\in\{0,1,2\}^V$ by the sets 
\begin{equation*}
    \begin{split}
        V_0(f) =& \{v_i\mid i\in [t], v\in V \}\cup \{a_i,b_i\mid i\in [t]\},\\
        V_1(f) =& \{v'\mid v\notin S\} ,\\
        V_2(f) =& \{x,y\} \cup \{v'\mid v\in S\}.
    \end{split}    
\end{equation*} 
 This implies $\weightfun(f)= \vert V\setminus S\vert+  4 + 2 \cdot\vert S\vert = \vert V\vert + 4 + \vert S\vert\leq k'.$
 The set $\{a_1,\ldots,a_t\} \cup \{ v' \mid f(v)\neq 2\}$ is only dominated by~$x$ and the vertices $b_1,\ldots,b_t$ are only dominated by~$y$. Assume there is a $v\in V$ and $i\in [t]$ such that $\vert N_{G'}(v_i)\cap V_2(f)\vert \neq 1$. By $N_{G'}(v_i)=\{u' \mid u\in N_G[v]\}$, this implies $\vert N_{G'}(v_i)\cap V_2(f)\vert = \vert N_G(v) \cap S\vert \neq 1$. This contradicts that $S$ is a perfect dominating set.
 
 Assume that $f\in \{0,1,2\}^{V'}$ is a minimum \prdf with $\weightfun(f)\leq k'$. 
 \begin{claim}
 For each $z\in \{v_i,a_i,b_i \mid i\in [t], v\in V \}$, it holds that $f(z)=0$.
 \end{claim}
 \begin{pfclaim}
 For the sake of contradiction, assume that there is some $v\in V$ and $i\in [t]$ with $f(v_i)= 2$. 
 If $v_i$ would not have a private neighbor~$u'$ with $f$-value~0, then $f'= f-\chi_{\{v_i\}}$ would be a smaller perfect Roman dominating function. This contradicts the minimality of~$f$. This implies 
 $\{x,y\}\cup \{z' \mid z\in V\}\subseteq N_{G'}(u')$ has no vertex with $f$-value~2. Thus, $N_{G'}(v_i)\cap V_2(f)=\emptyset$. Therefore, $f(v_j)\neq 0$ for each $j\in [t]$. Hence $\weightfun(f)\geq t + 1 > k'$, contradicting the choice of~$f$. 
 
 Assume there is some $v\in V$ and $i\in [t]$ with $f(v_i)= 1$.
 Furthermore, we can assume $\vert N_{G'}(v_i)\cap V_2(f)\vert \neq 1$, as otherwise, $f-\chi_{\{v_i\}}$ would be a smaller \prdf. As by construction $N_{G'}(v_i)=N_{G'}(v_j)$ for each $j\in [t]$, we find $\vert N_{G'}(v_i)\cap V_2(f)\vert = \vert N_{G'}(v_j) \cap V_2(f)\vert$. If $\vert N_{G'}(v_i)\cap V_2(f)\vert=0$, then $f(v_j)\neq 0$ holds for each $j\in [t]$, since $v_j$ is not dominated. Hence, $\weightfun(f)\geq t\cdot|V|\geq t>k'$, contradicting the choice of~$f$.
 Assume $\vert N_{G'}(v_i)\cap V_2(f)\vert\geq 2$. Thus, for each $j\in [t]$, $f(v_j)\neq 0$, as otherwise, $f$ is not perfect.
 This contradicts the construction as above. 

 We can discuss the claims concerning $a_i$ and $b_i$ along very similar lines.
 \end{pfclaim}
 
 Further, we know $f(x)=f(y)=2$. Define $S=\{v\mid f(v')=2\}$. As $f(v_i)=0$ for each $v\in V, i\in [t]$, $\vert N_{G'}(v_i)\cap V_2(f)\vert=1$. This implies that $\vert N_G[v]\cap S\vert=1$ for each $v\in V$. Therefore, $S$ is a perfect dominating set.
 
 Since $f(x)=f(y)=2$, $f(v')\geq1$ for all $v\in V$. This implies that $\weightfun(f)=\vert V\vert + \vert V_2(f)\vert +4=\vert V\vert + \vert S\vert +4\leq k'$. Hence, $\vert S \vert \leq k'- 4 - \vert V\vert = k$.
 \end{pf}

 \begin{theorem}
     $\weightfunction{f}-\PRD$ can be solved in $\FPT$ time on split graphs.
 \end{theorem}

 \begin{pf}
Let $G=(V,E)$ be a split graph with the vertex set partition $V = C\cup I$, where $C$ is a clique and $I$ an independent set. Further, let $f:V \to \{0,1,2\}$ be a minimum \prdf on $G$. Then $V_2(f) \cap I=\emptyset$ or $V_2(f) \cap C=\emptyset$. For $v\in V_2(f) \cap I$ and $c\in V_2(f) \cap C$, $f-\chi_{\{v\}}$ would be a \prdf with $\weightfunction{f-\chi_{\{v\}}} \leq \weightfunction{f}$. 

For $V_2(f)\cap C=\emptyset$, we just branch on each vertex in $I$ if gets the value~$1$ or~$2$. The vertices cannot get the value~$0$ as they cannot be dominated by a other vertex. After deciding this we can delete the vertex and the parameter goes down by either~$1$ or~$2$. This gives us the branching vector $(1,2)$. When we have branched on each vertex in $I$, then we assign $0$ to each $c\in C$ with $\vert V_2(f) \cap N(c)\vert = 1$ and $1$ otherwise (this is important to get the \prdf property). Now, we only need to check if $\weightfunction{f}$ is smaller than the given parameter value. 

Now, we consider the \prdfs $f\in \{0,1,2\}^V$ with $ V_2(f)\cap I =\emptyset$. As $C$ is a clique, if $\vert V_2(f) \cap C \vert \geq 2 $, then $C\subseteq V_1(f) \cup V_2(f)$. 
First we consider the \prdfs with $\vert V_2(f)\cap C\vert=1$. Since these functions are \urrdfs, this can be done in polynomial time. After this, we guess two vertices from $C$ and assign the value~$2$ to these vertices. From now on, we can branch on the remaining vertices in $C$ if these will be assigned to~$1$ or~$2$. Analogously to $V_2(f)\cap C=\emptyset$, then we assign $0$ to each $v\in I$ with $\vert V_2(f) \cap N(v)\vert = 1$ and $1$, otherwise. In this case, we also have to check if $\weightfunction{f}$ is smaller than the given parameter value.

Our algorithm runs in time $\Oh^*(\Phi^k)$, where $\Phi$ is the golden ratio and $k$ the parameter. 
 \end{pf}

 \subsection{Perfect Roman Domination on Cobipartite Graphs}
 
For the remaining section, we consider cobipartite graphs, which are the complementary graphs of bipartite graphs. These graphs can be also characterized as graph for which the vertex set can be partitioned into two cliques. On this graph class, \PRD\ is solvable polynomial-time. 

\begin{theorem}
 \PRD\ is polynomial-time solvable on cobipartite graphs.    
\end{theorem}

\begin{pf}
     Let $G=(V,E)$ be a cobipartite graph with the partition into the two cliques $C_1,C_2\subseteq V$. For $v\in V$ define the \prdf 
     $$g_v: V \to \{0,1,2\}, x\mapsto\begin{cases}
         0,& x\in N(v)\\
         1,& x\in V \setminus N[v]\\
         2,& x=v.
     \end{cases}$$
     Let $f \in \{0,1,2\}^V$ be a \prdf with $ \vert V_2(f) \cap  C_1\vert\geq 2 $ and $v\in C_2$.
     For each $u\in C_1\setminus V_2(f)$, $\vert V_2(f) \cap N(u)\vert \geq 2$. Hence $C_1\subseteq V_1(f) \cup V_2(f)$. Since $C_2\subseteq N[v] =V_0(g_v) \cup \{ v\}$, $\weightfunction{g_v}\leq \vert C_1 \vert +2 \leq \weightfunction{f}$. Symmetrically, $\weightfunction{g_v}\leq \vert C_1 \vert +2 \leq \weightfunction{f}$ for $v\in C_1$ and a \prdf $f$ with $ \vert V_2(f) \cap  C_2\vert\geq 2 $.
     Let $f$ be a minimum \prdf on $G$. We can assume $\vert V_2(f) \cap C_i\vert\leq 1$ for each $i\in\{ 1, 2\}$. This leaves only $\vert V\vert ^2 +\vert V\vert$ many possibilities. 
\end{pf}

\subsection{Relating to 2-Packings}
As mentioned before, Cabrera \emph{et al.}~\cite{CabPueRod2021} proved that \textsc{Unique Response Roman Domination} is \NP-complete on bipartite graphs. For cobipartite graphs, this problem is polynomial-time solvable. For this result, we use the idea of 2-packings. Let $G=(V,E)$ be a graph. A set $S\subseteq V$ is called \emph{2-packing} if the distance between two vertices in $S$ is at least 3. Targhi \emph{et al.}~\cite{TarRadVol2011} presented a proof for $u_r(G) = \min\{2\vert S \vert + \vert V(G) \setminus N_G[S]\vert \mid S \text{ is a 2-packing}  \}$. Analogously to this, we can prove that, for each graph $G=(V,E)$, there exists a bijection $\psi_G$ between all 2-packings of the graph and all \urrdfs. Here, for a 2-packing~$S$ and for $x\in V$, 
$$\psi_G(S)(x)=\begin{cases}
    0, & x \in N(S) \setminus S \\
    1, & x \in V \setminus N(S)\\
    2, & x \in S.
\end{cases}$$
This also yields that, for each \urrdf $f$, $V_2(f)$ is a 2-packing. This is the main idea of our algorithm.

\begin{lemma}
    \textsc{Unique Response Roman Domination} is polynomial-time solvable on cobipartite graphs. Furthermore,  there are at most $\frac{\vert V\vert^2}{4}$ many 2-packings, or \urrdfs, on $G=(V,E)$. 
\end{lemma}

\begin{pf}
    Let $G=(V,E)$ be a cobipartite graph, together with the partition into the two cliques $C_1,C_2\subseteq V$. Clearly, two vertices from the same clique cannot be in a 2-packing at the same time. Therefore, from each clique, there can be at most one vertex in the 2-packing. This implies that there are at most $\vert C_1\vert \cdot \vert C_2\vert$ many 2-packings, or \urrdfs. As $C_1,C_2$ is a partition of $V$, this leads to $\vert C_1\vert \cdot \vert  C_2\vert =\vert C_1\vert \cdot \vert V\setminus C_1\vert = \vert C_1\vert \cdot \left( \vert V\vert - \vert C_1\vert\right)$ many 2-packings. This expression is maximal if $\vert C_1\vert =\vert C_2\vert$, which implies that there are at most $\frac{\vert V\vert^2}{4}$ many 2-packings or \urrdfs on~$G$.
\end{pf}

The bound $\frac{\vert V\vert^2}{4}$ is tight: we get the number for the complement of a complete bipartite graph $K_{t,t}$ where both classes have the same size~$t$. This is interesting as there could be exponentially many \urrdfs\ even on connected split graphs, which is quite a related class of graphs. To this end, we only need to consider the split graph $G=(V,E)$ with $V\coloneqq C\cup I$, $C \coloneqq \{c_1,\ldots,c_t\}$, $I \coloneqq \{v_1,\ldots, v_{2t}\}$ and 
$$E \coloneqq \binom{C}{2}\cup \{\{c_i,v_{2i-1}\}, \{c_i,v_{2i}\} \mid i\in \{1,\ldots,t\}\}.$$
Clearly, $\vert V\vert = 3t$. By the arguments from \ref{cor:URRDF_P_on_split}, we know that for each \urrdf $f$ with $\vert V_2(f) \cap C \vert =1$, $\vert V_2(f) \vert = 1$. There are $t$ many such \urrdfs. Let $f$ be a \urrdf with $ V_2(f) \cap C = \emptyset$.  In this situation, for each $i\in \{1, \ldots, t\}$, there three ways to dominate $c_i$:
\begin{itemize}
    \item $f(c_i) = f(v_{2i-1}) = f(v_{2i}) = 1 $,
    \item $f(c_i) = f(v_{2i}) =0$ and $f(v_{2i-1}) = 2$,
    \item $f(c_i) = f(v_{2i-1}) =0$ and $f(v_{2i}) = 2$.
\end{itemize}
This implies that there are $t+3^t= \frac{\vert V \vert}{3} + \sqrt[3]{3}^{\vert V \vert}$ many \urrdf. $\Omega(\sqrt[3]{3}^{\vert V \vert})$ is even a tight bound for connected split graphs. This is the case as each \urrdf on a graph without isolated vertices is a minimal \rdf. Therefore, we could use the enumeration algorithm for minimal \rdfs on split graph from \cite{AbuFerMan2023} which runs in $\mathcal{O}(\sqrt[3]{3}^{\vert V \vert})$.  

\begin{remark}
It should be mentioned that the polynomial delay property of this algorithm  is not inherited for enumerating \urrdf, as not each minimal \rdf is a \urrdf. Nonetheless, we will present a sketch of a polynomial-delay  branching enumeration algorithm. For this purpose, we consider each 2-packing on a connected split graph and use the bijection between \urrdfs and 2-packings. The measure is for each vertex the same.

Let $G=(V,E)$ be a connected split graph with the partion $V=C\cup I$ where $C$ is a clique and $I$ is an independent set. With the arguments from above, we can first enumerate each of the $\vert C \vert$ many 2-packings $S$ with $S \cap C\neq \emptyset$. From now on, we branch on the vertices in~$I$ if they will be in~$S$ or not. If the vertex from~$I$ is not in $S$, then we delete this vertex. If we put a vertex from $v$ into $S$, then we delete, for each $u\in N(v)$, the whole set $\{u\}\cup \left( N(u)\cap I\right)$. Furthermore, we delete vertices from~$C$ which have no neighbors in~$I$.  We branch on the vertex $v \in I$ with the highest degree. If $\deg(v)\geq 3$, then by putting the vertex in~$S$ we delete~4 vertices. This leads to a branching vector $(4,1)$ (the branching value is below 1.3804). For $\deg(v)=2$, we have to make a case distinction in the analysis:
If $v$ is the only neighbor in~$I$ for each of the neighbors of~$v$, then in both cases these vertices will be deleted after this branch. This implies the branching vector $(3,3)$ (The branching value is $\sqrt[3]{2}<\sqrt[3]{3}$). Otherwise, at least 4 vertices will be deleted after putting~$v$ into~$S$. This would again lead to the branching vector $(4,1)$.

Therefore, we can assume that each vertex in~$I$ has exactly one neighbor. Let $c$ be this neighbor. If $\vert N(c)\cap I\vert \geq 3$, after putting $v$ into $S$, we would delete at least 4 vertices and the branching would be again $(4,1)$. For  $N(c) \cap =\{v\}$, we would delete $c$ in each case. This leads to the branching vector $(2,2)$ (the branching value is $\sqrt{2}\leq \sqrt[3]{3}$). The remaining case is   $\vert N(c)\cap I\vert =2$. This is considered in the $\Omega(\sqrt[3]{3}^{\vert V \vert})$ example above. This implies the claimed running time of $\Oh^*({\sqrt[3]{3}\,}^n)$.

This algorithm is a modification of the recursive backtracking algorithm of~\cite{Mar2015} and is polynomial delay with the same arguments. The algorithm of~\cite{Mar2015} is a general approach to enumerate all sets of $\mathcal{F}\subseteq 2^U$ for a universe $U$, where $\mathcal{F}$ fulfills the downward closure property: if $X\in \mathcal{F}$ and $Y \subseteq X$, then $Y\in\mathcal{F}$.
\end{remark}
\begin{corollary}
    There exists an algorithm which enumerates all \urrdfs of a connected split graph of order~$n$ in time $\Oh^*(\sqrt[3]{3}^{n})$ with polynomial delay. Furthermore, there are connected split graphs of order $n$ that have at least $\Omega({\sqrt[3]{3}\,}^n)$ many \urrdfs.
\end{corollary}

\section{Enumerating All Unique Response Roman Dominating Functions 
}\label{sec:Enum_urrdf}

In this section, we will enumerate \urrdfs on graphs without isolated vertices. The reason for this restriction is that there are two choices for a \urrdf to dominate an isolated vertex (either 1 or 2). This would result in $2^n$ many \urrdfs. In this case, an algorithm which decides for each vertex if it assigns the value~$2$ or not (at the end each vertex $v$ without a value-2 vertex in its closed neighborhood will be assigned~1, and it will be assigned~0, otherwise) would have perfect running time. 
Furthermore, the value~$2$ on an isolated vertex is no good idea, as in the \textsc{Unique Response Roman Domination} problem, we try to minimize the value of the function. In this case, each \urrdf that assigns a~$2$ to an isolated vertex would give rise to a smaller \urrdf which dominates the remaining vertices in the same way. Therefore, as we are going to enumerate all \urrdfs in this section, not only the minimal ones, we are considering graphs without isolated vertices in the following.

Recall that Junosza-Szaniawski and  and Rzążewski~\cite{JunRza2012} provided an enumeration algorithm  for 2-packings on connected graphs of order~$n$ which runs in time $\Oh(1.5399^n)$. Remember that there exists a bijection between all 2-packings and all \urrdfs of a graph. Even if this is an interesting algorithm for enumerating all \urrdfs on connected graphs, there are worse cases on graphs without isolated vertices:

On $P_2=(\{v,u\}, \{\{v,u\}\})$, there are the following three possible \urrdfs: $2\cdot \chi_{\{v\}},2\cdot \chi_{\{u\}}$ and $\chi_{\{v,u\}}$.

\begin{corollary}
There are graphs of order $n$ without isolates that have at least ${\sqrt{3}\,}^n$ many minimal \urrdf.
\end{corollary}
We will also need the next observation in the following.

The remaining section will be used to show the following theorem that basically proves that the given simple example is optimal.

\begin{theorem}\label{thm:minimal-urrdf-enumeration}
There is a polynomial-space algorithm that enumerates all \urrdfs of a given graph (without isolated vertices) of order $n$ with polynomial delay and in time $\mathcal{O}^*({\sqrt{3}\,}^n)$.
\end{theorem}
For the proof, we will construct a branching algorithm. To do this, we need the sets $\Active, V_0,V_1,V_2,\overline{V_2}$. In $V_0,V_1,V_2$, we find the vertices with the respective values 0, 1, 2. Hence, $V_2$ has to be a 2-packing. $\overline{V_2}$ contains the vertices which will not be~$2$ but it is not clear if they are $0$ or $1$ ($\overline{V_2} \cap (V_0 \cup V_1)= \emptyset$). $\Active$ contains vertices which are completely undecided so far, which reflects the situation at the start of the algorithm. For the analysis, we use the measure $\mu=\vert \Active \vert + \omega \cdot\vert \overline{V_2}\vert$. Notice that at the very beginning, $V=A$, so that then $\mu=|V|$.

For the polynomial delay part of the proof, we will consider the search tree representation of a run of a branching algorithm on a graph~$G$. This is a rooted tree where all edges are orientated away from the root. Each node represents a quintuple $(\Active,\overline{V_2}, V_0, V_1, V_2)$ of pairwise disjoint sets whose union is~$V$. There is an edge from one node to another one in this tree if the second node can be produced in one of the cases of the next branching step. Reduction rules can be executed in polynomial time and we hence stay within the same node of the search tree.
 
\begin{redrule}\label{rr:urRDF_Neighbored_V2_vertices}
If there are $v,u\in V_2$, $u\neq v$, with $N_G[v]\cap N_G[u] \neq \emptyset$, then skip this branch.
\end{redrule}

In other words, we have detected a leaf in the search tree in which no solution is output. This reduction rule is  sound as $V_2$ would not be a 2-packing anymore.

\begin{redrule}\label{rr:urDF_vertex_without_privateneighbor}
If there is a vertex $v\in \Active$ with $N_G(v)\cap(\overline{V_2} \cup \Active) = \emptyset$, then put $v$ into~$\overline{V_2}$.
\end{redrule}

\begin{lemma}
\autoref{rr:urDF_vertex_without_privateneighbor} is sound.
\end{lemma}

\begin{pf}
Let $f$ be a \urrdf on $G$ with $V_2(f)\cap (V_2 \cup \overline{V_2})= V_2$, $V_0 \subseteq V_0(f)$ and $V_1 \subseteq V_1(f)$. Since we assigned 0 to a vertex only if it has a neighbor in $V_2$ and $N_G(u)\cap V_2(f)$ is empty for each $u\in V_1(f)\cup V_2(f)$, $v\in V_2(f)$ would contradict the properties of \urrdfs. 
\end{pf}

\begin{redrule}\label{rr:urRDF_notV2_not_dominated}
If there is a vertex $v\in \overline{V_2}$ with $N_G[v] \cap \Active =\emptyset$, then put $v$ into~$V_1$.
\end{redrule}

\begin{lemma}
\autoref{rr:urRDF_notV2_not_dominated} is sound.
\end{lemma}

\begin{pf}
Since $N_G[v]\cap V_2(f)$ will be empty for a \urrdf $f$ with $V_2 \subseteq V_2(f)$ and $\overline{V_2} \cap V_2(f)=\emptyset$, $f(v)$ must be~1. 
\end{pf}

\begin{brarule}\label{br:urRDF_twoActiveNeighbors}
Let $v\in \Active$ with $\vert N_G(v) \cap \Active\vert \geq 2$. Then branch as follows:
\begin{enumerate}
    \item Put $v$ in $V_2$ and all vertices of $N_G(v)$ in $V_0$.
    \item Put $v$ in $\overline{V_2}$.
\end{enumerate}
\end{brarule}

\begin{lemma}
The branching is a complete case distinction. Moreover, it leads at least to the following branching vector:
$(3,1 - \omega)\, .$
\end{lemma}

\begin{pf}
If a vertex is in $V_2$, then each neighbor has to be in $V_0$. Therefore, this is a complete case distinction. In the first case, we put three vertices from $\Active$ into $V_2$ or $V_0$. Hence, the measure decreases by~$3$. The last case is decreasing the measure by $ 1 - \omega $, since we only move $v$ from $\Active$ to $\overline{V_2}$.
\end{pf}

\begin{brarule}\label{br:urRDF_pedant_noActive2Neighbors}
Let $v\in \Active$ with $\{u\} = N_G(v) \cap A$ and $\{v\} = N_G(u) \cap A$. Then branch as follows:
\begin{enumerate}
    \item Put $v$ in $V_2$ and $N_G(v)$ in $V_0$.
    \item Put $u$ in $V_2$ and $N_G(u)$ in $V_0$.
    \item Put $v,u$ in $V_1$.
\end{enumerate}
\end{brarule}

\begin{lemma}
The branching is a complete case distinction. Moreover, it leads at least to the following branching vector:
$(2,2,2)\, .$
\end{lemma}

\begin{pf}
In this case $v$ and $u$ cannot be in $V_2$ at the same time. If $v,u\in \overline{V_2}$, then these vertices cannot be dominated, as $N_G(v) \cap (V_2 \cup A) = N_G(u) \cap (V_2 \cup A) = \emptyset$. Since the values of $v,u$ are fixed after this branch, the measure decreases by (at least)~$2$ in each case.
\end{pf}

\begin{brarule}\label{br:urRDF_JustV2Neighbors}
Let $v\in \Active$ with $N_G(v) \subseteq \overline{V_2}$. Then branch as follows:
\begin{enumerate}
    \item Put $v$ in $V_2$ and all vertices of $N_G(v)$ in $V_0$.
    \item Put $v$ in $V_1$.
\end{enumerate}
\end{brarule}

\begin{lemma}
The branching is a complete case distinction. Moreover, it leads at least to the following branching vector:
$(1+ \omega,1)\, .$
\end{lemma}

\begin{pf}
Because of \autoref{rr:urDF_vertex_without_privateneighbor}, we know that $v$ has at least one neighbor in $\overline{V_2}$. If $v$ is assigned to~$2$, then its neighbors have to be in $V_0$ (this decreases the measure by at least $1 + \omega$). $v\in \overline{V_2}$ triggers \autoref{rr:urRDF_notV2_not_dominated}, which implies $v\in V_1$. This reduces the measure by~$1$.  
\end{pf}

\begin{proof}[\autoref{thm:minimal-urrdf-enumeration}]
First we will show that the algorithm covers each case. \autoref{br:urRDF_twoActiveNeighbors} implies that each vector in $\Active$ has at most one neighbor in $\Active$. By \autoref{br:urRDF_pedant_noActive2Neighbors}, there is no edge between two vertices in~$\Active$. If there is a vertex in $\Active$ with a neighbor in $\overline{V_2}$, then this triggers \autoref{br:urRDF_JustV2Neighbors}. For the remaining vertices in~$\Active$, we use \autoref{rr:urDF_vertex_without_privateneighbor}.
If a vertex in $\overline{V_2}$ has no neighbor in~$\Active$, we use \autoref{rr:urRDF_notV2_not_dominated}. Hence, the branching algorithm handles each case for $V \nsubseteq V_0\cup V_1\cup V_2$. As each branching is complete the branching algorithm returns all \urrdfs of $G$. The running time of the algorithm follows by \autoref{tab:branching-vectors-urRDF}.

By an inductive argument, it can be shown that after putting a vertex into $V_0,\, V_1$ or $V_2$ we will never remove it from these sets again. Further, after putting a vertex into $\overline{V_2}$, it will only go into $V_0$ or~$V_1$. Since we decide for $\Active$ vertices to be in either $V_2$ or $\overline{V_2}$ (together with the use of \autoref{rr:urRDF_notV2_not_dominated}), this implies that if the algorithm returns two \urrdfs, then they will not be the same.   

\begin{claim}
 Let $(\Active,\overline{V_2},V_0,V_1,V_2)$ be a constellation which is represented by a node~$v$ in the branching tree. Then either we can use \autoref{rr:urRDF_Neighbored_V2_vertices}, or there is a constellation $(\emptyset,\emptyset,V_0',V_1',V_2')$ which is represented by a leaf of the branching tree below~$v$ such that $f\in \{0,1,2\}^V$ is a \urrdf with $V_i(f)\coloneqq V'_i$ for $i\in\{0,1,2\}$, is a \urrdf.
\end{claim}
\begin{pfclaim}
In each branching rule, the algorithm puts a vertex into $V_0$ \iffl the vertex has a neighbor in $V_2$. Further, the neighbors of $V_2$ vertices are all in~$V_0$.  The algorithm puts a vertex into $V_1$ only if we use \autoref{rr:urRDF_notV2_not_dominated}. As the algorithm never puts a vertex from $\overline{V_2}$ into~$V_2$,  $V_1 \cap N_G(V_2)$ is empty. 

Define $g : V \to \{0, 1, 2\}$ with $V_0(g):= V_0$, $V_1(g) = \Active \cup \overline{V_2} \cup V_1 $ and $V_2(g) = V_2$. We get $g$ by using the last case of each branching rule on $(\Active,\overline{V_2},V_0,V_1,V_2)$. By our observations from above, $g$ is a \urrdf \iffl $\vert N_G(v)\cap V_2\vert =1$ holds for each $v\in V_0(g)=V_0$. If this would not hold, then we could use \autoref{rr:urRDF_Neighbored_V2_vertices}.  
\end{pfclaim}

As we only consider each vertex at most once in a branching, we have to test \autoref{rr:urRDF_Neighbored_V2_vertices} at most $\vert V\vert$ times. This test runs in polynomial time. Therefore, the branching algorithm runs with polynomial delay. 
\end{proof}

\begin{table}[tb]
    \centering
    \begin{tabular}{l|l|l}
        Branching Rule \# & Branching vector & Branching number\\\hline
        \autoref{br:urRDF_twoActiveNeighbors} & $(3,1-\omega)$ & 1.7229\\
        \autoref{br:urRDF_pedant_noActive2Neighbors} & $(2,2,2)$ & $\sqrt{3}$\\
        \autoref{br:urRDF_JustV2Neighbors} & $(1 + \omega, 1)$ & 1.7218
\end{tabular}
    \caption{Collection of all branching vectors; the branching numbers are displayed for the different cases with $\omega=0.6$.
    }
    \label{tab:branching-vectors-urRDF}
\end{table}

\section{Extension Perfect Roman Domination}\label{sec:ExtPrd}

In this section, we will consider \textsc{Extension Perfect Roman Domination}. More precisely, we will first provide a combinatorial result for deciding if a function is a (pointwise) minimal perfect Roman dominating function.

\begin{lemma}\label{lem:conditionsPRDF}
Let $G=(V,E)$ be a graph. A function $f:V\rightarrow \{0,1,2\}$ is a minimal \prdf \iffl the following conditions are true:
\begin{enumerate}
    \item $\forall v\in V_0(f): \: \vert N_G(v) \cap V_2(f) \vert = 1$,\label{con:PRDFV0}
    \item $\forall v\in V_1(f): \: \vert N_G(v) \cap V_2(f) \vert  \neq 1$,\label{con:PRDFV1}
    \item $\forall v\in V_2(f): \:\vert N_G(v) \cap V_0(f) \vert \neq 0$.  \label{con:PRDFV2}
\end{enumerate}
\end{lemma}

\begin{pf}
Let $f$ be a minimal \prdf. This implies the first condition. If there exists a $v\in V_1(f)$ with $\vert N_G(v)\cap V_2(f)\vert = 1$, then $f-\chi_{\{v\} }$ is also a \prdf, cf. \autoref{obs:private-URRD}. 
For each $v\in V_2(f)$, $\vert N_G(v) \cap V_0(f) \vert \neq 0$, as otherwise $f-\chi_{\{v\} }$ is also a \prdf.

Now assume $f$ is a function that fulfills the three conditions. By the first condition, we know $f$ is a \prdf. Let $g$ be a minimal \prdf with $g\leq f$. Therefore, $V_0(f)\subseteq V_0(g)$ and $V_2(g)\subseteq V_2(f)$ hold. Assume there exists a $v\in V$ with $g(v) < f(v)=2$.
By the third condition, there exists some $u\in N_G(v)\cap V_0(f)$. Because of condition~1, $v$ is the only neighbor of~$u$ with value~2. This implies that $N_G(u) \cap V_2(g)\subseteq N_G(u) \cap (V_2(g) \setminus \{v\})=\emptyset$. This contradicts the construction of~$g$ as a \prdf.  Therefore, $V_2(g)=V_2(f)$.  If there were a $v\in V$ with $g(v)=0<1=f(v)$, then this contradicts that~$g$ is a \prdf, since $\vert N_G(v)\cap V_2(g)\vert = \vert N_G(v)\cap V_2(f) \vert \neq 1$. 
\end{pf}

\begin{remark}
    This lemma provides that the function $\mathds{1} \colon V \to \{0,1,2\}$ is the maximum minimal \prdf with respect to $\weightfun$ for each graph $G=(V,E)$, similar to maximum minimal \rdfs as discussed in~\cite{CheHHHM2016}. This is the case as if there exists a vertex $v\in V$ of value~$2$ then this vertex needs a neighbor $u$ with value~$0$. Since $u$ has exactly one neighbor in $V_2(f)$, $\weightfunction{f} \leq \vert V\vert$ for each \prdf $f$. 
\end{remark}

 We will use this lemma to show some some hardness results for \textsc{Extension Perfect Roman Domination}. 

\centerline{\fbox{\begin{minipage}{.99\textwidth}
\textbf{Problem name: }\textsc{Extension Perfect Roman Domination}\\
\textbf{Given: } A graph $G=(V,E)$ and $f\in \{0,1,2\}^V$.\\
\textbf{Question: } Is there a minimal \prdf $g$ on $G$ with $f \leq g$?\end{minipage}
}}

 \begin{remark}
It should be mentioned that there are graphs $G=(V,E)$ such that the set of \prdf{}s on $G$ is not closed under $\leq$. This means that there exists a \prdf $f \in \{0,1,2\}^V$ and a $g \in \{0,1,2\}^V$ with $f \leq g$ such that $g$ is no \prdf.  One example for this is the 4-cycle $$C_4=(\{v_1,v_2,v_3,v_4\}, \{\{v_1,v_2\},\{v_2,v_3\},\{v_3,v_4\},\{v_4,v_1\}\})$$ with the two functions
$$
\begin{tabular}{cccc}
         $f(v_1)=0$, & $f(v_2)=2$, & $f(v_3)=0$, & $f(v_4)=1$, \\
         $g(v_1)=0$, & $g(v_2)=2$, & $g(v_3)=0$, & $g(v_4)=2$.
\end{tabular}
$$
This contradicts the idea of extension problems as they are defined in \cite{CasFGMS2021}, because this violates the so-called upward closedness condition, but this condition is a matter of debate even in~\cite{CasFGMS2021}.
\end{remark}

\subsection{\W{1}-Hardness}

In this subsection, we will show \W{1}-hardness of  \EPRD if parameterized by the weight of the pre-solution. As the reduction will also be polynomial, this will also imply $\NP$-hardness of the underlying unparameterized problem. For the $\NP$-membership, we simple guess the value of each vertex (at most $\vert V\vert$ steps) and then check the conditions of \autoref{lem:conditionsPRDF}.
  For the hardness result, we use \textsc{Irredundant Set}. A set $I\subseteq V$ is called \emph{irredundant} if each vertex in $I$ has a private neighbor.

\centerline{\fbox{\begin{minipage}{.99\textwidth}
\textbf{Problem name: }\textsc{Irredundant Set}\\
\textbf{Given: } A graph $G=(V,E)$ and $k\in \mathbb{N}$\\
\textbf{Question: } Is there an irredundant set $I$ on $G$ with $\vert I \vert = k$?\end{minipage}
}}

\noindent
In \cite{DowFelRam2000} it is shown that \textsc{Irredundant Set}, parameterized by $k$, is \W{1}-complete.

\begin{theorem}\label{thm:ExtPrdf_NP}
\EPRD is \NP-complete.
$k$-\EPRD is \W{1}-hard (even on bipartite graphs).
\end{theorem}

\begin{pf}
For the \NP-membership, we simple guess the value of each vertex and check the conditions of \autoref{lem:conditionsPRDF}.

Let $G=(V,E)$ be a graph and $k\in \mathbb{N}$. Define for each $i\in [k+1]$, $V_i=\{v_i \mid v\in V\}$ and $G'=(V',E')$ with  
\begin{equation*}
    \begin{split}
        V' =\,& \{a,b,c,d\} \cup \{u_1, \ldots, u_k\} \cup \bigcup_{i=1}^{k+1} V_i,\\
        E' =\,& \{\{a,b\},\{c,d\}\}\\ \cup\,& \{ \{a,u_i\}, \{v_i,u_i\}, \{v_i,c\}, \{v_i, w_{k+1}\} \mid v\in V, w\in N_G[v],i\in [k]\}.
    \end{split}
\end{equation*}
We also need $f\in \{0,1,2\}^{V'}$ with $
        V_0(f)=\{b,d\} \cup \bigcup_{i=1}^{k+1} V_i, V_1(f)=\{ u_1,\ldots, u_k\}$ and $V_2(f)=\{a,c\}.$
This implies $\weightfun(f)=k+4$.
$G'$ is a bipartite graph, as we can partition its vertex set into two independent sets $I_1=\{a,d\}\cup \{v_i\mid v\in V, i\in [k]\}$ and $I_2=\{b,c\}\cup \{u_i\mid i\in [k]\}\cup\{v_{k+1}\mid v\in V\}$.

Let $I=\{w^1,\ldots,w^k\} \subseteq V$ be an irredundant set with $\vert I\vert = k$. Define $g\in \{0,1,2\}^{V'}$ with
\begin{equation*}
    \begin{split}
        V_0(g)=&\,\{b,d\} \cup \{v_{k+1}\mid v \in V: \vert N_G[v] \cap I\vert = 1\}\cup \{ v_i \mid i \in [k], v\in V\setminus \{w^i\}\}, \\
        V_1(g)=&\,\{ u_1,\ldots, u_k\}  \cup \{v_{k+1}\mid v \in V : \vert N_G(v) \cap  I\vert \neq 1\}, \\
        V_2(g)=&\,\{a,c\}\cup \{w_1^1,\ldots, w_k^k\}.
    \end{split}
\end{equation*}

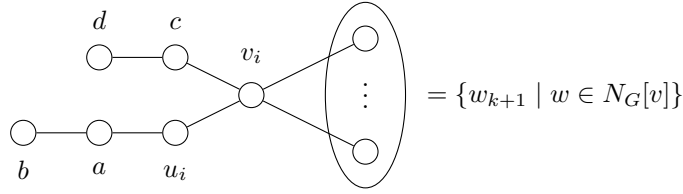
\begin{figure}[bt]
    \centering
    	
	\begin{tikzpicture}[transform shape]
		      \tikzset{every node/.style={ fill = white,circle,minimum size=0.3cm}}
            \node  at (4,0){$=\{w_{k+1}\mid w\in N_G[v]\}$};

			\node[draw,label={above:$v_i$}] (vi) at (0,0) {};
			\node[draw,label={below:$u_i$}] (ui) at (-1,-0.5) {};
			\node[draw,label={below:$a$}] (a) at (-2,-0.5) {};
			\node[draw,label={below:$b$}] (b) at (-3,-0.5) {};
			\node[draw,label={above:$c$}] (c) at (-1,0.5) {};
			\node[draw,label={above:$d$}] (d) at (-2,0.5) {};

			\node[draw] (vi1) at (1.5,-0.75) {};
			\node[draw] (vi2) at (1.5,0.75) {};


            \node at (1.5,0.1) {\vdots};
   
            \draw (1.5,0) ellipse (15pt and 35pt);
            \path (vi) edge[-] (ui);	
            \path (vi) edge[-] (c);	
            \path (b) edge[-] (a);
            \path (ui) edge[-] (a);	
            \path (d) edge[-] (c);	
            \path (vi) edge[-] (vi1);	
            \path (vi) edge[-] (vi2);	
        \end{tikzpicture}

        \label{fig:defall_TJ_PSpace}
    \caption{Construction for \autoref{thm:ExtPrdf_NP}, for each $v\in V$ and $i\in [k]$.}
\end{figure}

Clearly, $f\leq g$ holds. The vertex~$b$ is a private neighbor of~$a$, and $d$ is one of~$c$.  For each $i\in [k]$, $g(w^i_i)=2$ implies $\vert N_{G'}(u_i)\cap V_2(g) \vert\geq 2$. Since $I$ is irredundant, for each $w^i\in I$, there is a private neighbor $p^i$ in~$G$ and hence,  for each $w^i_i$, there is a private neighbor $p^i_{k+1}\in \{v_{k+1}\mid v\in V\}$ in $G'$. As $g$ clearly is a Roman domination function, it is hence also a \prdf by \autoref{obs:private-URRD}. Furthermore, for all $i\in [k]$ and $v\in  V \setminus \{w^i\}$, $\{c\}= N_{G'}(v_i)\cap V_2(g)$ holds. By the construction of~$g$, all vertices in $v_{k+1}$ with $v\in V$ have the correct value for $g$ to be a minimal \prdf.

Assume we have a minimal \prdf $g$ on $G'$ with $f\leq g$. For all $i \in [k]$, $N_{G'}(u_i)=\{a\} \cup \{v_i\mid v\in V\} \subseteq \{a\} \cup N_{G'}(c)$. As $f(c)=g(c)=2$, this implies that $g(u_i)=1$ and that $V_i\cap V_2(g)$ is not empty by \autoref{lem:conditionsPRDF}. Let $w^i_i$ be such a vertex for each $i\in [k]$. Define $I = \{w^1, \ldots, w^k\}\subseteq V$. As for each $i\in [k]$, all vertices in $V_i$ are already dominated by~$c$, the private neighbor of $w_i^i$ has to be $p^i_{k+1}\in V_{k+1}$ and hence, $p^i$ is a private neighbor of $w^i$ in~$G$. Therefore, $I$ is an irredundant set with $\vert I\vert =k$.
\end{pf}
This result implies that it is unlikely to have an \FPT{} algorithm for $k$-\EPRD (unless $\W{1}=\FPT$).
Nonetheless, we can provide an \XP algorithm below.

As $\vert V_2(f)\vert =2$ and $\vert V_1(f)\vert=k$, we get the following corollary.

\begin{corollary}
    $\vert V_1(f)\vert$-\EPRD is \W{1}-hard and $\vert V_2(f)\vert$-\EPRD is \paraNP-hard. 
\end{corollary}

\subsection{\XP-Membership}

As we ruled out membership in \FPT{} under standard complexity assumptions in the previous subsection, the next best algorithmic fact one could hope for is membership in \XP, a class that is also dubbed `poor man's \FPT'.
In the following, we will provide an explicit \XP-algorithm, allowing a running time of $\Oh^*(n^k)$ on graphs of order~$n$.
Notice that in the next subsection, we even prove membership in \W{1}.
From this membership, one can also deduce membership in \XP, but without having an implementable algorithm in hands. Further, the \XP algorithm implies some theoretical results which we need for the \W{1} membership proof. 
As testified in~\cite{Sch2022} for quite related problems (also confer the discussions in~\cite{FerMan2023}), also \XP-algorithms of the proposed form could be very helpful in practical implementations.

\begin{lemma}\label{lem:goal_xp_algo}
    Let $G=(V,E)$ be a graph and $f\in \{0,1,2\}^V$ such that for each $u \in V_1(f)$,  $\vert N_G(u) \cap V_2(f)\vert \neq 1$ and for each $v\in V_2(f)$, there exists a private neighbor in $V_0(f)$ with respect to $G$ and $V_2(f)$. Then there exists a minimal \prdf $g\in \{0,1,2\}^V$ on $G$ with $f\leq g$ and $V_2(f)=V_2(g)$.
\end{lemma}

\begin{pf}
    Let $A = \{v\in V_0(f)\mid \vert N_G(v) \cap V_2(f) \vert \neq 1 \}$. Define $g=f+\chi_A$. Clearly, $f\leq g$ and $V_2(f)=V_2(g)$. This leaves to show that $g$ is a minimal \prdf. 

    By definition of $A$, for each $v \in V_{0}(g)=V_0(f)\setminus A$, $1 = \vert N_G(v) \cap V_2(f)\vert= \vert N_G(v) \cap V_2(g) \vert$. Further for all $v \in V_1(g)=V_1(f) \cup A$,  $1 \neq \vert N_G(v) \cap V_2(f)\vert$. Let $v\in V_2(g)$. By the requirement on $V_2(f)=V_2(g)$, for $v\in V_2(g)$ there exists a private neighbor $u \in V_0(f)$. Thus, $\vert N_G(u) \cap V_2(f)\vert = 1$ and $u\in V_0(f)\setminus A= V_0(g)$. Hence $ N_G(v) \cap V_0(f) \neq \emptyset$ for all $v\in V_2(f)$. In total $g$ is a minimal Roman dominating function.
\end{pf}

The proof gives also an algorithm to compute the minimal \prdf from the given function $f$. We only have to compute $A\subseteq V_0(f)$ as specified in the proof, which can be done in polynomial time.

\begin{lemma}\label{lem:goal_extprdf_member}
    Let $G=(V,E)$ be graph and $f\in \{0,1,2\}^V$. Then there exists a minimal perfect Roman dominating function $g\in \{0,1,2\}$ with $f\leq g$ and $V_2(f) =V_2(g)$ \iffl each $v\in V_2(f)$ has a private neighbor in $V_0(f)$ (with respect to $G$ and $V_2(f)$) and $\vert N_G(u) \cap V_2(f)\vert \neq 1$ for each $u\in V_1(f)$.
\end{lemma}
\begin{pf}
    The if-part follows directly by \autoref{lem:goal_xp_algo}.

    For the only if part, assume there exists a minimal perfect Roman dominating function $g$ on $G$ with $f \leq g$ and $V_2(f)=V_2(g)$. Hence, $V_0(g) \subseteq V_0 (f)$ and $V_1(f) \subseteq V_1(g)$. Since $g$ is a minimal perfect Roman dominating function, Conditions \ref{con:PRDFV0} and \ref{con:PRDFV2} hold. This implies that each $v\in V_2(g)=V_2(f)$ has a private neighbor in $V_0(g)\subseteq V_0(f)$. For each $u\in V_1(g)$, $\vert N_G(u)\cap V_2(f)\vert=\vert N_G(u)\cap V_2(g)\vert\neq 1 $. As $V_1(f) \subseteq V_1(g)$, this also holds for all $u\in V_1(f)$.
\end{pf}

\noindent
Now we can prove that \autoref{alg:ExtPRDF} is an \XP algorithm.
\begin{algorithm}
\caption{Solving instances of \textsc{ExtPRD}}\label{alg:ExtPRDF}
\begin{algorithmic}[1]
\Procedure{ExtPRD Solver}{$G,f$}\newline
 \textbf{Input:} A graph $G=\left(V,E\right)$ and a function $f\colon V\to \lbrace0,1,2\rbrace$.\newline
 \textbf{Output:} Is there a minimal \prdf $\widetilde{f}$ with $f\leq \widetilde{f}$?
 \For{$v\in V_2(f)$}
    \State $x\coloneqq 0$
    \For{$w \in V_0(f) \cap V_2(f)$}
        \If{$\vert N(w) \cap V_2(f) \vert = 1$}
            $x=1$
        \EndIf
    \EndFor
    \If{$x=0$}
        \Return \no
    \EndIf
 \EndFor
 \For{$v\in V_1(f)$}
    \If{$\vert N(v) \cap V_2(f) \vert =1$} 
        \For{$u\in N(v)\setminus V_2(f)$}
            \State $ f_u \coloneqq f$
            \State Set $f_u(u)=2$
            \If{\textsc{ExtPRD Solver}$(G,f_u)$}
            \Return \yes
            \EndIf
        \EndFor
        \State \Return \no
    \EndIf
 \EndFor
 \State \Return \yes
\EndProcedure
\end{algorithmic}
\end{algorithm}

\begin{theorem}
    \autoref{alg:ExtPRDF} is an $\XP$ algorithm for $k$-$\textsc{Extension}$ $\textsc{Perfect}$ $\textsc{Roman}$ $\textsc{Domination}$ for $k\in \{\weightfunction{f}, \vert V_1(f) \vert\}$. Let $(G,f)$ with $G=(V,E)$ and $f\in \{0,1,2\}^V$ be an instance of $\textsc{Extension}$ $\textsc{Perfect}$ $\textsc{Roman}$ $\textsc{Domination}$. If the algorithm returns a minimal perfect Roman dominating function $g$, then $\vert V_2(g)\vert \leq \weightfunction{f}$.
\end{theorem}

\begin{pf}
Let $G=(V,E)$ be graph and $f\in \{0,1,2\}^V$. Since $\vert V_1\vert \leq \weightfunction{f}$, we only need to consider $\vert V_1(f)\vert$ as parameterization. At first we explain the algorithm. The idea is modify $f$ such that it fulfills the conditions of \autoref{lem:goal_xp_algo}. Therefore, we look if each $v\in V_2(f)$ has a private neighbor in $V_0(f)$. If this is not the case, then there exists no minimal perfect Roman dominating function bigger than $f$. If each $v\in V_2(f)$ has a private neighbor in $V_0(f)$, we look if there exists a $u\in V_1(f)$ with $\vert N_G(u) \cap V_2(f)\vert =1$. If this is not the case, then $f$ itself already fulfills the condition of \autoref{lem:goal_xp_algo}. Otherwise, we go through all neighboring vertices $v\in N_G(u)\setminus V_2(f)$ and try the algorithm with 
    $$f_{v}\colon V \to \{0,1,2\}, \, x \mapsto \begin{cases}
        2, & x=v\\
        f(x), & x\neq v
    \end{cases},$$
\emph{i.e.}, $f_v=f+(2-f(v))\cdot \chi_{\{v\}}$.
Since we only return \yes (or a minimal perfect Roman dominating function) if $f$ verifies the condition of \autoref{lem:goal_xp_algo} and we only increase $f$, there exists a minimal perfect Roman dominating function $g\in \{0,1,2\}$ with $f\leq g$ if we return \yes (or a minimal perfect Roman dominating function). 

Assume there exists a minimal \prdf, say $g$, with $f \leq g$. This implies $V_2(f)\subseteq V_2(g)$ and $V_0(g) \subseteq V_0(f)$. We prove that the algorithm will return a minimal perfect Roman dominating function by an induction argument on $\weightfun(g-f)$. 

If $\weightfun(g-f)=0$, $f=g$ holds. Then each vertex $V_2(f)$ has a private neighbor in $V_0(f)$. Otherwise, it would contradict \autoref{lem:conditionsPRDF}, Condition~\ref{con:PRDFV2}. Further, for $f=g$ there exists no vertex $v \in V_1(f)$ with $\vert N_G(v) \cap V_2(f)\vert =1$ because of Condition~\ref{con:PRDFV1} of \autoref{lem:conditionsPRDF}.

Assume $\weightfun(g-f)=1$. Then there exists exactly one vertex in $v\in V$ with $f(x)\neq g(v)$. More precisely, $f(v)=g(v)-1$. First, we consider the case $f(v)=0$. This implies that $V_2(f)=V_2(g)$. By \autoref{lem:goal_extprdf_member}, $f$ fulfills the conditions of \autoref{lem:goal_xp_algo}.
Now assume $f(v)=1$. This implies $V_0(f)=V_0(g)$ and $V_2(g)=V_2(f)\cup \{v\}$. Therefore, for each $u \in V_2(f)\subseteq V_2(g)$, there exists a private neighbor in $N_G(u)\cap V_0(g)=N_G(u)\cap V_0(f)$. If each $u\in V_1(f)$ fulfills $\vert N_G(u) \cap V_2(f) \vert \neq 1 $, then our algorithm would make use of \autoref{lem:goal_xp_algo}. Assume there exists a $u\in V_1(f)$ with $\{w\}= N_G(u) \cap V_2(f)$. If $u \notin N_G(v)$, then $\vert N_G(u) \cap V_2(g)\vert=1 $ would contradict the minimality of $g$. Therefore, $v\in N_G(u)$. Since $\vert N_G(u) \cap V_2(g)\vert = 1$ holds, our algorithm would try, for each $x \in N_G[u]\setminus \{w\}$, if there exists a minimal perfect Roman dominating function $h\in \{0,1,2\}^V$ such that $f_x \leq h$. Hence, our algorithm would also consider $f_v=g$ (see above).  

Let $\weightfun(g-f)>2$ and and assume that $f$ does not fulfill the conditions of  \autoref{lem:goal_xp_algo}. Hence, $f\neq g$. Recall $V_2(f)\subseteq V_2(g)$ and $V_0(g)\subseteq V_0(f)$. Therefore, for each $v\in V_2(f)\subseteq V_2(g)$, there exists a $u\in N_G(v) \cap V_0(g) \subseteq N_G(v) \cap V_2(f)$ with $ N_G(u) \cap (V_2(f)\setminus \{v\}) \subseteq N_G(u) \cap (V_2(g)\setminus \{v\}) = \emptyset$. This implies that each $v\in V_2(f)$ has a private neighbor in $V_0(f)$. Since the algorithm does not return a minimal  perfect Roman dominating function, there exists a $u\in V_1(f)$ with $\{v\}= N_G(u) \cap V_2(f)$. As $g$ is minimal, there must exist a $w\in (N_G(v)\cap V_2(g))\setminus \{u\}$. Therefore, we consider $f_x$ for each  $x\in N_G[u]\setminus \{v\}$ (so also for $w$). By induction and $\weightfun(g-f_w) < \weightfun(g-f)\leq \weightfun(g-f_w) + 2$, it follows that we find a minimal perfect Roman dominating function $f'$ on $G$ with $f\leq f'$. Hence, the algorithm runs correctly.  

Now, we consider the running time of the algorithm.
Checking if each vertex in $V_2(f)$ has a private neighbor in $V_0(f)$ can be done in polynomial time (in time $\mathcal{O}(\vert V\vert^3)$ with a naive algorithm). Further, we can test in polynomial time if there exists a $u\in V_1(f)$ with $\vert N_G(u) \cap V_2(f)\vert =1$ (in time $\mathcal{O}(\vert V\vert^2)$ with a naive algorithm). If this is the case, then we go through $w\in N_G[u] \setminus V_2(f)$ and run the algorithm on on $f_w$. Clearly $V_0(f_w)\subseteq V_0(f)$ and $V_1(f_w) \subseteq V_1(f)$ and $ V_2(f)\cup \{w\}=V_2(f_w)$. Together with $\vert N_G(u) \cap V_2(f_w)\vert > 1$, this implies that $\vert N_G(u) \cap V_2(h)\vert \neq  1$ will hold for all $h\in \{0,1,2\}^V$ with $f_w \leq h$. Thus, we add only one vertex to $V_2(f)$ per vertex in $V_1(f)$. As we  never add a vertex to $V_1(f)$ (unless we use \autoref{lem:goal_xp_algo}, but then we already know that there is a solution), the recursion tree has at most $\vert V_1(f) \vert \leq  k$ many nodes between the root and a leaf. This also proves the bound on the size of $V_2(h)$ for a solution $h$ returned by the algorithm. As there are at most $\vert V\vert $ choices for $w$, we call the recursive function at most $n^k$ times. Therefore, it is an \XP algorithm. 
\end{pf}

\noindent
The proof of the last theorem implies the following corollary.

\begin{corollary}\label{cor:minimal_prdf_max_size_V_2}
    Let $G=(V,E)$ be a graph and $f\in \{0,1,2\}^V$ with $k\coloneqq \weightfunction{f}$. If there exists a minimal perfect Roman domination function $h$ on $G$ with $f\leq h$, then there exists a minimal perfect Roman domination function $g$ with $\vert V_2(g)\setminus V_2(f)\vert \leq \vert V_1(f)\vert$,  $\vert V_2(g)\vert \leq k$ and $f\leq g$.
\end{corollary}

The very existence of an \XP-algorithm with respect to $\vert V_1(f)\vert$ is interesting, as the very similarly looking parameterized problem $\vert V_1(f) \vert $-$\textsc{Extension}$  $\textsc{Roman}$ $\textsc{Hitting Set}$ is $\paraNP$-hard (see \cite{FerMan2023}).
Moreover, we need this corollary in order to prove \W{1}-membership in the next subsection, so that we cannot derive \XP-membership without the considerations of this subsection.

\subsection{\W{1}-Membership}

We do not only provide a hardness result (presented in the previous subsection) but also a complete classification of $\weightfunction{f}$-\textsc{Extension Perfect Roman Domination}.

\begin{lemma}\label{lem:idea_extprdf_member}
Let $G=(V,E)$ be a graph and $f\in \{0,1,2\}^V$ with $k\coloneqq \weightfunction{f}$. There exists a minimal \prdf $g$ on $G$ with $f \leq g$ \iffl there exists a $V'\subseteq V$ with $\vert V' \vert \leq k$ and $V_2 \subseteq V'$ such that each $v\in V'$ has a private neighbor in $V_0(f)\setminus V'$ with respect to $G$ and $\vert N_G(u) \cap V' \vert \neq 1$ for each $u \in V_1(f) \setminus V'$.  
\end{lemma}

\begin{pf}
Let $g$ be a minimal \prdf on $G$ with $f \leq g$. By \autoref{cor:minimal_prdf_max_size_V_2}, there exists a minimal \prdf $g'$ on $G$ with $f \leq g'$ and $\vert  V_2(g') \vert \leq k$. As $f\leq g'$, $V_2(f)\subseteq V'\coloneqq V_2(g')$, $V_0(g') \subseteq V_0(f) \setminus V'$ and $V_1(f)\setminus V'\subseteq V_1(g')$. Since $g'$ is minimal \prdf, each $v\in V'$ has a neighbor in $V_0(g')\subseteq V_0(f) \setminus V'$. By the definition of \prdf, this neighbor is private. For $u\in V_1(f)\setminus V'\subseteq V_1(g')$, $\vert N_G(u) \cap  V'\vert \neq 1$. Therefore $V'$ fulfills the conditions.

Let $V'$ be a set that verifies the condition. Then $f':V\to \{ 0,1,2\}$ with $V_0(f')=V_0(f)\setminus V'$, $V_1(f')=V_1(f)\setminus V'$ and $V_2(f')=V'$ fulfills the conditions of \ref{lem:goal_extprdf_member}. Hence, there exists a minimal \prdf $g$ on $G$ with $f\leq f'\leq g$.
\end{pf}

We will use this result to show \W{1}-membership by a reduction with \textsc{Short Non-Deterministic Turing Machine Computation}.

\centerline{\fbox{\begin{minipage}{.99\textwidth}
\textbf{Problem name: }\textsc{Short Non-Deterministic Turing Machine Computation}\\
\textbf{Given: } A nondeterministic one-tape Turing machine TM, a word $w$ and $k\in \mathbb{N}$\\
\textbf{Parameter: } $k$\\
\textbf{Question: } Does TM accept $w$ in at most $k$ steps?\end{minipage}
}}

\begin{theorem}\label{thm:W1_complete_ExtPRDF}
    $k$-\textsc{Extension Perfect Roman Domination} is \W{1}-complete.
\end{theorem}

\begin{pf}
    We only need to prove \W{1}-membership. Therefore, let $G=(V,E)$ be a graph and $f\in \{0,1,2\}^V$ be a function with $V_1(f) \coloneqq \{u^1,\ldots,u^{\ell}\}$ and $V_2(f)=\{v^1,\ldots,v^{\ell'}\}$ ($\ell + 2\ell' =k$ and $0 \leq \ell, \ell'$).
    
    The idea of our nondeterministic Turing machine (NTM for short) is to guess at most $k$ new vertices in $V_2(g)$ for a \prdf $g$ with $f\leq g$. After this we check if the vertices satisfy the conditions of \autoref{lem:goal_extprdf_member}. For the private neighborhood condition we guess for each vertex which vertex is the private neighbor and check if this vertex is the private neighbor and if the vertex is in $V_0(f)$. For the second condition we go for each $u\in V_1(f)$ through the vertices and count the number of neighbors up to two. 

    Now we will describe how the NTM works. The input alphabet is given by the vertex set. Then, $\Gamma \coloneqq \{b\} \cup V\cup \{v_i\mid v\in V, i\in [k]\}$ ($\vert \Gamma \vert = \vert V \vert \cdot (k + 1) $) denotes the tape alphabet for which $b$ is the blank symbol. Define the set of states by 
    \begin{equation*}
        \begin{split}
            Q \coloneqq{} & \{q_f\}\cup \{s_1, \ldots, s_{\ell'}\} \cup \{q_{\text{fill},\ell'+1}, \ldots,q_{\text{fill},k+1}\}\cup \{q_w^{j,i} \mid w \in V,\,i,j\in [k], \, i < j\} \\
        {}\cup{} & \{q_L^{j,i} \mid i,j\in [k], i < j\} \cup \{q_j^0,q_j^1,q_{L,j}\mid j\in [\ell]\}.
        \end{split}
    \end{equation*}
    There is only the final state $q_f$ and $s_1$ is the start state. The number of states is at most $k^2\cdot (\vert V\vert +1)+\Oh(k)$.
The input word is just $\ldots,b, v^1, \ldots, v^{\ell'},b,\ldots$\,\,. Now we will present all transitions:
    \begin{enumerate}
        \item $((s_i,v^i),(s_{i+1},v^i_i,R))$ for all $v\in V_2(f)$ and $i\in [\ell'-1]$,\footnote{Let us briefly explain how we write down the transitions of a Turing machine using this example. This transition can be activated if the TM is in state $s_i$ and currently reads the symbol $v^i$. It can then move to state $s_{i+1}$, replace the symbol $v^i$ by $v^i_i$ and then move to the right.}\label{rel:start}
        \item $((s_{\ell'},v^{\ell'}),(q_{\text{fill}, \ell' + 1}, v_{\ell'}^{\ell'}, R))$ \label{rel:start_to_fill}
        \item $((q_{\text{fill},i},b),(q_{\text{fill},i+1},v_{i},R))$ for all $i\in [k]\setminus [\ell']$ and $v\in V$,\label{rel:fill}
        \item $((q_{\text{fill},i},b),(q_{L}^{i-1},b,L))$ for all $i\in [k+1]\setminus \{ 1 \}$,\label{rel:end_fill}
        

        \item \textcolor{black}{$((q_{L}^i,v_p),(q_{L}^{i},v_p,L))$ for all $v\in V$ and $i,p\in [k]$ with $p \leq i$,}\label{rel:first_L}
        
        \item 
        $((q_{L}^i,b),(q_{w}^{1,i},b,R))$ for all $w\in V$ and $i\in [k]$,\label{rel:privacy_begin}
        \item $((q_{w}^{t,i},v_t),(q_{w}^{t,i},v_t,R))$ for all $\{v,w\}\in E$ and $t,i\in [k]$ with $t\leq i$,\label{rel:privacy_neighbor}
        \item $((q_{w}^{t,i},v_j),(q_{w}^{t,i},v_j,R))$ for all $v,w\in V$ and $i,j,t\in [k]$ with $v \notin N_G[w]$ and $j,t\leq i$, \label{rel:privacy_private}
        \item $((q_{w}^{t,i},b),(q_{L}^{t,i},b,L))$ for all $w\in V\setminus V_1(f)$ and $i,t\in [k]$ with and $t\leq i$, \label{rel:privacy_to_left}
        \item $((q_{L}^{t,i},v_j),(q_{L}^{t,i},v_j,L))$ for all $v\in V$ and $i,j,t\in [k]$ with $j,t\leq i$,\label{rel:privacy_left}
        \item $((q_{L}^{t,i},b),(q_{w}^{t+1,i},b,R))$ for all $w\in V$ and $i,t\in [k]$ with $t < i$,\label{rel:privacy_new_round}
        \item $((q_{L}^{i,i},b),(q_{1}^{0},b,R))$ for all $w\in V$ and $i \in [k]$,\label{rel:privacy_end}
        \item $((q_{t}^{0},v_j),(q_{t}^{0},v_j,R))$ for all $v\in V$, $t \in [\ell]$ and $j\in [k]$ with $v\notin N_G[u^t]$,\label{rel:V1_0_no_neighbor}
        \item $((q_{t}^{0},v_j),(q_{t}^{1},v_j,R))$ for all $v\in V$, $t \in [\ell]$ and $j\in [k]$ with $\{ v, u^t\}\in E$,\label{rel:V1_0_neighbor}
        \item $((q_{t}^{0},v_j),(q_{L,t},v_j,L))$ for all $v\in V$, $t \in [\ell]$ and $j\in [k]$ with $v= u^t$,\label{rel:V1_0_same_vertex}
        \item $((q_{t}^{0},b),(q_{L,t},b,L))$ for all $t \in [\ell]$,\label{rel:V1_0_end}
        \item $((q_{t}^{1},v_j),(q_{t}^{1},v_j,R))$ for all $v\in V$, $t \in [\ell]$ and $j \in [k]$ with $v \notin N_G[u^t]$,\label{rel:V1_1_no_neighbor}
        \item $((q_{t}^{1},v_j),(q_{L,t},v_j,L))$ for all $v\in V$, $t \in [\ell]$ and $j \in [k]$ with $v \in N_G[u^t]$, \label{rel:V1_1_closed_neighbor}
        \item $((q_{L,t},v_j),(q_{L,t},v_j,L))$ for all $v \in V$, $t \in [\ell]$ and $j \in [k]$,\label{rel:V1_left}
        \item $((q_{L,t},b),(q_{t+1}^0,b,R))$ for all $t \in [\ell]$,\label{rel:V1_new_round}
        \item $((q_{L,\ell},b),(q_{f},b,L))$.\label{rel:end}
    \end{enumerate}
    For the proof we will divide a run of the NTM into three phases and prove claims for each phase separately.
    The first phase is from the beginning to the (only) use of  the transition \ref{rel:privacy_begin}. In this phase, we guess $V_2(g)$ of our would-be solution~$g$.
    
    At the beginning of the run, we are in the state $s_1$ and at the leftmost position of the tape with a non-blank symbol ($v^1$ is in this position). By an easy inductive argument, we can prove that after $\ell'$ steps, $\ldots,b,v_1^1,\ldots,v^{\ell'}_{\ell'},b,\ldots$ is written on the tape and the tape head is to the right of $v_{\ell'}^{\ell'}$ and the TM is in the state $q_{\text{fill},\ell'+1}$. Keep in mind that the states $s_1,\ldots,s_{\ell'}$ occur only in the transitions~\ref{rel:start}, being mentioned in increasing order. Therefore, we will never get into these states again.    
    
    The next step is the first nondeterministic step. Here the Turing machine guesses, if it uses a transition of \ref{rel:end_fill} or uses for a $v\in V$ the transition $$((q_{\text{fill},\ell'},b),(q_{\text{fill},\ell'+1},v_{\ell'+1},R))\,.$$ 
    If the Turing machine uses the second case, \emph{i.e.}, when a new (encoded) vertex has been written on the tape, then it has to guess again if it moves into the state $q_{L}^{i}$ or if it writes a new vertex on the tape ($ \ell' \leq i \leq k$). After at most $k+1$ steps, we have to use a transition from~\ref{rel:end_fill}, as for $q_{\text{fill},k+1}$ there is only one transition. 
    
    Assume that after $i$ steps (with $ \ell' \leq i \leq k$) we use a transition from~\ref{rel:end_fill}. Then we are in the state $q_{L}^i$ and $\ldots,b,v_1^1,\ldots,v_i^i,b\ldots$ is written on the tape. More precisely, the initial string $v^1\cdots v^{\ell'}$ that was the input of the TM has been converted into the prefix $v^1_1\cdots v^{\ell'}_{\ell'}$, while the suffix $v^{\ell'+1}_{\ell'+1}\cdots v_i^i$ is the part that was nondeterministically guessed.
    Since the NTM is never going in to the states $s_1,\ldots,s_{\ell'}$ again and \ref{rel:start_to_fill} and \ref{rel:fill} are the only transitions with a $q_{\text{fill},j}$ on the right side for $j\in [k]\setminus [\ell']$, the Turing machine never goes back into these states.
    From now on, the next $i$ steps are deterministic. By an easy inductive argument again, the tape head moves to the position to the left of $v_1^1$. At this point, the first phase ends. This phase needed $2i+1 \leq 2k+1$ steps. As we never use the transitions from  \ref{rel:start} to \ref{rel:privacy_begin} again, the symbols on the tape do not change anymore, \emph{i.e.}, during the remaining run,  $\ldots,b,v_1^1,\ldots,v_i^i,b\ldots$ will stay on the tape, where $v^1,\ldots,v^i\in V$ are not necessary different vertices. Since $q_{\text{fill},j}$ had to choose any vertex from $V_2(f)$, $V'\coloneqq\{v^1,\ldots, v^i\}$ could be any set of size at most $i$ with $V_2(f) \subseteq V'$. 

    The next phase deals with the privacy condition of the vertices on the tape. This will also ensure that no vertex is written twice on the tape. This phase will end with the use of a transition from~\ref{rel:privacy_end}. The transitions in \ref{rel:privacy_begin} and \ref{rel:privacy_new_round} are nondeterministic, as the Turing machine guesses the vertex~$w$. Let $t\in [i]$ and $w\in V_1(f)$. The idea of the state $q_{w}^{t,i}$ is to check if $w$ is the private neighbor of $v^t$ (and $v^t \neq w$) with respect to $V'$. Let $j\in [i]$. If $j=t$, then there is a deterministic step \iffl $w\in N_G(v_t)$ (see \ref{rel:privacy_neighbor}). Otherwise, the Turing machine stops (in a non-final state), as there is no transition the NTM can use. For $j\neq t$ the Turing machine will only stop (in a non-final state) \iffl $w\in N_G[v_j]$. Otherwise, it performs a deterministic step and goes on with the next symbol (vertex) $v_{j+1}$ (see~\ref{rel:privacy_private}). In other words, the Turing machine stops in the state $q_{w}^{t,i}$ \iffl $w$ is not a private neighbor of $v_t$ and $v_t\neq w$. This can be proven by induction again. If the Turing machine made $i$ steps in the state $q_{w}^{t,i}$, the head is at the end of the word $v_1^1,\ldots,v^i_i$ and goes back with  the state $q^{t,i}_L$ to the beginning (see~\ref{rel:privacy_to_left} and~\ref{rel:privacy_left}). Then, the state changes to $q_{w}^{t+1,i}$ for $t\in [i-1]$ (see~\ref{rel:privacy_new_round}) or it goes into the new state $q_{1}^0$ for $t=i$ (see~\ref{rel:privacy_end}). Therefore, after $2i^2$ steps in this phase, the Turing machine is still running \iffl each vertex $V'$ has a private neighbor in $V\setminus (V_1(f) \cap V')$. This ends the second phase. 

    The third phase  will check if each $u\in V_1(f)\setminus V'$ verifies $\vert N_G(u)\cap V'\vert \neq 1$. Then \autoref{lem:idea_extprdf_member} implies that there exists a minimal \prdf $g$ on $G$ with $f \leq g$. The idea of the state $q_{t}^z$ for $t\in [\ell]$ and $z\in \{0,1\}$, is that the index~$z$ counts how many neighbors of $u^t$ we have seen so far.
    
    Let $t\in [\ell]$. Assume we are in the state $q_{t}^0$ and at the current position on the tape, there is $v_j^j$ for $j\in [i]$. If $v^j=u^t$, then $u^t\in V'\cap V_1(f)$. Thus, we do not need to consider $u^t$ anymore and we switch into the state $q_{L,t}$ (see~\ref{rel:V1_0_same_vertex}). If $v_j\in N_G(u)$, the Turing machine has seen a neighbor and goes on with the state $q_t^1$ (see~\ref{rel:V1_0_neighbor}). Otherwise, we just go to the right, staying in the same state (see~\ref{rel:V1_0_no_neighbor}). In the case when the tape head is at the right end of the word in the state $q^0_{t}$, this implies that $u^t$ has no neighbor in~$V'$. Therefore, the NTM switches the state to $q_{L,t}$ (see~\ref{rel:V1_left}). 
    
    Assume we are  in the state $q_t^1$. Keep in mind that the NTM only goes into the state $q_t^1$ \iffl the vertex under the current head position is in the neighborhood of~$u^t$. For $v^j\in N_G[u^t]$, either $u^t=v^j\in V'\cap V_1(f)$ or $\vert N_G(u^t) \cap V'\vert >1$. In both cases, we need not consider $u^t$ anymore and the Turing machine can switch into the state $q_{L,t}$ (see \ref{rel:V1_1_closed_neighbor}). If $v_j\notin N_G(u^t)$, the NTM only goes to the right on the tape (see~\ref{rel:V1_1_no_neighbor}).
    By an inductive argument, we can show that the NTM ends in the state $q_t^i$ on a cell containing~$b$ \iffl $u^t$ has exactly one neighbor in $V'$. In this case, the Turing machine would stop in a non-final state. 
    This implies that the NTM uses the transition~\ref{rel:end} after at most $2 \ell \cdot i$ steps in the last phase \iffl for each $t\in [\ell]$, $\vert N_G(u^t) \cap V' \vert \neq 1$. 

    In total, the Turing machine reaches $q_f$ in at most $2i + 1 +2i^2+ 2 \ell \cdot i\leq k'\coloneqq 4k^2+2k+1$ steps \iffl there exists a vertex set $V'\supseteq V_2(f)$ with $\vert V' \vert \leq k$ such that each $v \in V'$ has a private neighbor in  $V_0(f)\setminus V'$ and for each $u\in V_1(f)\setminus V'$, $\vert N_G(u)\vert \neq 1$. By \autoref{lem:idea_extprdf_member}, the Turing machine ends in a final state (after at most $k'$ steps) \iffl there exists a minimal \prdf $g$ on $G$ with $f \leq g$.
\end{pf}

\subsection{$\vert V_0(f)\vert$-\textsc{Extension Perfect Roman Domination}}

The main goal of this subsection is to prove the \W{2}-completeness of this parameterized problem. 
We start with some small observations.

\begin{lemma}
    $\vert V_0(f) \cup V_1(f)\vert$-$\textsc{Extension Perfect Roman Domination},\\  \weightfunction{2-f}$-$\textsc{Extension Perfect Roman Domination}\in \FPT$.
\end{lemma}
The proof works analogously to Theorem 55 of \cite{FerMan2023}.
\begin{proof}
    For the proof we go trough all possible functions $g\in \{0,1,2\}^V$ with $f \leq g$ and check if $g$ is a minimal \prdf (such a check runs in polynomial time). Let $v\in V$. If $f(v)=2$, then $g(v)=2$. For $v\in V_1(f)$, there are two choices for $g(v)$. If $f(v)=0$ then $g(v) \in \{0,1,2\}$. Since   $\vert V_0(f) \cup V_1(f) \vert \leq \weightfunction{2-f}$ and for each vertex in $V_0(f) \cup V_1(f)$ there are at most 3 choices, there up to $3^k$ possibilities to check for $k \in \{ \vert V_0(f) \cup V_1(f) \vert, \weightfunction{2-f}\}$. Hence, there is a \FPT algorithm for both parameterizations.  
\end{proof}

For $\weightfunction{2-f} $ as parameterization are even $2^{\weightfunction{2-f}}$ many possibilities. For the details take a look into \cite{FerMan2023}.

\begin{lemma}\label{lem:V2gleqV0f}
    Let $G=(V,E)$ be a graph and $f\in \{0,1,2\}^V$ a function. For a minimal \prdf $g$ on $G$ with $f \leq g$, $\vert V_2(g)\vert \leq \vert V_0(f)\vert$.
\end{lemma}

\begin{pf}
    As mentioned before, $V_0(g) \subseteq V_0(f)$. Further, $\vert N_G(v) \cap V_2(g) \vert = 1$ for each $v\in V_0(g)$, by \autoref{lem:conditionsPRDF}. Therefore, there exists a function $\phi: V_0(g) \to V_2(f)$, which maps a vertex to its unique neighbor in $V_2(f)$. By \autoref{lem:conditionsPRDF}, each vertex in $V_2(g)$ has at least one neighbor in $V_0(g)$. Thus, $\phi$ is surjective. Hence, $\vert V_2(g) \vert \leq \vert V_0(g) \vert\leq \vert V_0(f) \vert $. 
\end{pf}

From this lemma we can we can provide a simple \XP algorithm with the parameterization $\vert V_1(f) \vert$. 
We know that if there exists a minimal \prdf $g$ with $f\leq g$ then $\vert V_2(g)\vert \leq \vert V_0(f)\vert$. Therefore, we guess the at most $\vert V_0(f) \vert$ many vertices in $V_2(g)$ and use \autoref{lem:goal_extprdf_member} on the new function.

Nonetheless, \autoref{alg:ExtPRDF} is also an \XP algorithm with respect to the parameterization $\vert V_0(f) \vert$. The running time result follows as on one path of the branching tree we can only add $\vert V_0(f)\vert $ vertices to $V_2(f)$. Otherwise, not each vertex in $V_2(f)$ will have private in $V_0(f)$. Therefore, the depth of the branching tree is at most $\vert V_0(f)\vert$. 

For the membership, we use a reduction using the problem \textsc{Short Blind Non-Deterministic Multi-Tape Turing Machine Computation} which was introduced by Cattanéo and Perdrix in \cite{CatPer2014}. In that paper, they have also shown \W{2}-completeness of this Turing machine problem. The difference between a \emph{blind} multi-tape  nondeterministic  and a normal multi-tape  nondeterministic Turing machine is that the transitions can be independent of the symbols that are in the cells under the current head positions, \emph{i.e.}, the Turing machine may, but need not read the cell contents, and in this sense, it may be blind.\footnote{Possibly, \emph{oblivious} would have been a better term for this property, but we stick to the notion \emph{blind} as introduced in the mentioned paper.} 

\centerline{\fbox{\begin{minipage}{.99\textwidth}
\textbf{Problem name: }\textsc{Short Blind Non-Deterministic Multi-Tape Turing Machine Computation}\\
\textbf{Given: } A nondeterministic multi-tape Turing machine TM, a word $w$ and $k\in \mathbb{N}$\\
\textbf{Parameter: } $k$\\
\textbf{Question: } Does TM accept $w$ in at most $k$ steps?\end{minipage}
}}
    
\begin{theorem}
    $\vert V_0(f)\vert$-$\textsc{Extension Perfect Roman Domination} \in \W{2}$. 
\end{theorem}
\begin{pf}
    For the membership, we only sketch the proof, as most parts are analogous to the proof of \autoref{thm:W1_complete_ExtPRDF}. 
    In this proof, we have $\vert V_1(f)\vert + 1$ many tapes and add the symbol $\#$ to our work alphabet. On the first tape, the set~$V'$ will be enumerated (as in proof of \autoref{thm:W1_complete_ExtPRDF}, so that  $V_2(f) \subseteq V'$ and hopefully $V'=V_2(g)$ for a perfect rdf~$g$ that extends~$f$). Each remaining
    tape will represent a vertex in $V_1(f) \coloneqq \{u^1,\dots,u^{\ell}\}$. At the beginning, $\ldots,b,\#,b,\ldots$ is on each of these tapes and the head is on the $b$-occurrence immediately to the right of~$\#$. 

    In the first $2i + 1 +2i^2$ ($i$ is the cardinality of $V'$) steps, we simulate the NTM of the proof of  \autoref{thm:W1_complete_ExtPRDF} on the first tape. The other tapes stay the same. Hence, before we start the third phase, we know that the first tape contains a list of vertices (without repetitions) that meet the privacy condition. 
    The third phase is different. We go once again through the word on the first tape. Let $v_j^j$ for $j\in [i]$ be the symbol on the current cell. Then proceed for the $(t + 1)^{\text{st}}$ tape  (for $t\in [\vert V_1(f)\vert]$) as follows:
    \begin{itemize}
        \item If $v^j = u^t$, then write  two $\#$  on the $(t + 1)^{\text{st}}$  tape, moving to the right after writing~$\#$.
        \item If $v^j\in N_G(u^t)$, then write one $\#$  on the $(t + 1)^{\text{st}}$  tape and go one step to the right.
        \item If $ v^j\notin N_G[u^t]$, then do nothing on the $(t + 1)^{\text{st}}$  tape. 
    \end{itemize}

    When we are gone through the first tape, we (blindly) move on each tape (but the first tape) left  twice. $b$ is now in a current cell of these tapes \iffl the corresponding $u^t$ has exactly one neighbor in $V'$ and is not in $V'$. So by \autoref{lem:idea_extprdf_member}, we go into the final state \iffl each head is on a $\#$.

    Altogether, the described NTM would make at most  $2i + 1 +2i^2+i+2$ many steps. As $i=|V_2(g)|$, the claim follows with \autoref{lem:idea_extprdf_member}.
\end{pf}

For proving \W{2}-hardness, we use $k$-\textsc{Multicolored Dominating Set} which is known to be \W{2}-complete, see~\cite{LacPfa2012}.

\centerline{\fbox{\begin{minipage}{.99\textwidth}
\textbf{Problem name: }$k$-\textsc{Multicolored Dominating Set}\\
\textbf{Given: } A graph $G=(V,E)$, $k\in \mathbb{N}$ and a partition $W_1,\ldots,W_k$ of $V$\\
\textbf{Parameter: } $k$\\
\textbf{Question: } Is there a dominating set $D \subseteq V$ with $\vert W_i \cap D \vert = 1$ for each $i\in [k]$?\end{minipage}
}}

\begin{theorem}\label{thm:ExtPrdf_W2hard}
    $\vert V_0(f)\vert$-\textsc{Extension Perfect Roman Domination} is \W{2}-complete even on bipartite graphs.
\end{theorem}
\begin{figure}[bt]
\centering
    	
\begin{subfigure}[t]{.51\textwidth}
    \centering
    	
	\begin{tikzpicture}[transform shape]
		      \tikzset{every node/.style={ fill = white,circle,minimum size=0.3cm}}
            \node  at (3.6,0)[rectangle,label={[rotate=90]above:$\{u_1\mid u\in N[v]\}$}]{};
            \node [label={above:$=$}] at (1.8,-0.5){};

			\node[draw,label={below:$v_2$}] (v2) at (-0.5,0) {};
			\node[draw,label={below:$a$}] (a) at (-1.5,0) {};
			\node[draw,label={below:$b$}] (b) at (-2.5,0) {};
			\node[draw] (u1) at (1,0.75) {};
			\node[draw] (u'1) at (1,-0.75) {};
            \node at (1,0.15) {\vdots};
            \draw (1,0) ellipse (15pt and 35pt);			
            \path (u1) edge[-] (v2);
			\path (u'1) edge[-] (v2);
			\path (a) edge[-] (v2);
			\path (b) edge[-] (a);
        \end{tikzpicture}

    \subcaption{$v\in V$}
\end{subfigure}
\begin{subfigure}[t]{.43\textwidth}
    \centering
    	
	\begin{tikzpicture}[transform shape]
			\tikzset{every node/.style={ fill = white,circle,minimum size=0.3cm}}
            \node  at (-1,0)[rectangle,label={[rotate=90]above:$\{u_2\mid u\in N[v]\}$}]{};
            \node [label={above:$=$}] at (-1.8,-0.5){};

			\node[draw,label={below:$v_1$}] (v1) at (0.5,0) {};
			\node[draw,label={below:$x_t$}] (xt) at (1.5,0) {};
			\node[draw] (u2) at (-1,0.75) {};
			\node[draw] (u'2) at (-1,-0.75) {};
            \node at (-1,0.15) {\vdots};
            \draw (-1,0) ellipse (15pt and 35pt);			
            \path (u2) edge[-] (v1);
			\path (u'2) edge[-] (v1);
			\path (xt) edge[-] (v1);
        \end{tikzpicture}

    \subcaption{$t\in [k], v\in W_t$}
    \end{subfigure}
    \caption{Construction for \autoref{thm:ExtPrdf_W2hard}}
 \label{fig:ExtPrdf_W2hard}
\end{figure}    
\begin{pf}
    Let $G=(V,E)$ be a graph with the vertex set partition $W_1, \ldots, W_k$. Define $X \coloneqq \{x_1, \ldots, x_k\}$ and $\widetilde{G} = (\widetilde{V},\widetilde{E})$  with $a,b\notin V$,
    \begin{equation*}
        \begin{split}
            \widetilde{V} \coloneqq& \{v_1,v_2 \mid v\in V\}\cup X\cup \{a,b\}\\
            \widetilde{E} \coloneqq& \{\{a,b\}\} \cup \{\{a,v_2\} \mid v\in V \}  \cup \{\{v_1,x_j\}\mid v\in W_j\} \cup \\
            & \{\{v_1,u_2\} \mid v,u \in V, u\in N_G[v]\}.
        \end{split}
    \end{equation*}

The graph is also visualized in \autoref{fig:ExtPrdf_W2hard}. $\widetilde{G}$ is bipartite with two color classes $A\coloneqq V_1\cup \{a\}$ and $B \coloneqq X\cup V_2 \cup \{b\}$. To make it easier to verify for the reader, the vertices of $A$ are mentioned first in the definition of $\widetilde{E}$. Further, define $f : \widetilde{V} \to \{0,1,2\}$ with $V_0(f) \coloneqq X\cup \{b\}, \, V_1(f) \coloneqq \{v_1,v_2 \mid v\in V\}$ and $V_2(f)\coloneqq \{a\}$. Clearly, this is even a polynomial-time reduction and $\vert V_0(f) \vert = k + 1$.

Let $D$ be a dominating set of $G$ with $ \vert W_i\cap D \vert = 1$ for each $i\in [k]$. For $i\in [k]$, $u^i$ denotes the unique vertex in $W_i \cap D$. Define $U=\{u^1_1, \ldots, u^k_1\}\subseteq V_1$ and $g\in \{0,1,2\}^V$ such that $V_0(g)\coloneqq V_0(f)$, $V_1(g) \coloneqq V_1(f)\setminus U$ and $V_2(g) \coloneqq \{a\} \cup U$.

By the use of \ref{lem:conditionsPRDF}, we will show that $g$ is a minimal \prdf. First of all, $N_G(a)\cap V_0(g)=\{b\}$ and $N_G(b)\cap V_2(g)=\{a\}$. Furthermore, for each $i\in [k]$, $N_G(u^i_1)\cap V_0(g) = \{ x_i \}$ and $N_G(x_i)\cap V_2(g) = \{ u^i_1 \}$. This implies the first and last condition of \autoref{lem:conditionsPRDF}. Let $v\in V$. If $v_1\in V_1(g)$, then $N_G(v_1)\cap V_2(g)=\emptyset$. As $D$ is a dominating set, $D \cap N_G[v]$ is not empty. This implies that $\vert N_G(v_2)\cap V_2(g)\vert > 1$. Thus, $g$ is a minimal \prdf. 
    
Let $g\in \{ 0, 1, 2\}^V$ be a minimal \prdf on $\widetilde{G}$ with $f \leq g$. Since $N_G(v_2)\cap V_0(g) \subseteq N_G(v_2)\cap V_0(f) = \emptyset$ for all $v\in V$, $\{v_2\mid v\in V\} \subseteq V_1(g)$. $V_0(f)\cap N_G(a)=\{b\}$ implies $g(b)=0$. Define  $W'_i \coloneqq \{v_1\mid v\in W_i\}$ for $i\in [k]$. As $N_G(w_1)\cap V_0(f)=\{x_i\}$ holds for all $i\in [k]$ and $w_1 \in W'_i$, $\vert V_2(g) \cap W_i\vert \leq 1$ for all $i\in [k]$. Define $D \coloneqq \{w \in V \mid \exists i\in [k]: \: \{ w_1 \} =V_2(g)  \cap W'_i$. Hence, $\vert D \cap W_i\vert \leq 1$ for each $i\in [k]$. Let $v\in V$. Since $g(v_2)=1$ and $a \in N_G(v_2)\cap V_2(g)$, $N_G(v_2)\cap (V_2(g) \setminus \{a\})$ is not empty. $N_G(v_2) \setminus \{a\} \subseteq \{v_1\mid v\in V\}$ implies  $N_G[v]\cap D'\neq \emptyset$. Therefore, $D'$ is a dominating set. By adding an arbitrary vertex from $W_i$ with $W'_i\cap V_2(g)=\emptyset$ to $D'$, we get a solution to the \textsc{Colored Dominating Set} instance.
\end{pf}

\section{Minimal Perfect Roman Domination and Minimal Roman Domination}\label{sec:Enum_minPrd}

In this section, we will take a look at the connection between minimal Roman dominating functions and minimal perfect Roman dominating functions. 
To this end, we will use the following theorem from~\cite{AbuFerMan2022b}.

\begin{theorem}\label{t_property_min_rdf}
Let $G=\left(V,E\right)$ be a graph, $f: \: V \rightarrow \lbrace 0,1,2\rbrace$ be a function and let $G'\coloneqq G\left[ V_0\left(f\right)\cup V_2\left(f\right)\right]$. Then, $f$ is a minimal Roman dominating function if and only if the following conditions hold:
\begin{enumerate}
\item$N_G\left[V_2\left(f\right)\right]\cap V_1\left(f\right)=\emptyset$,\label{con_1_2}
\item $\forall v\in V_2\left(f\right) :\: P_{G',V_2\left(f\right)}\left( v \right) \nsubseteq \lbrace v\rbrace$, and \label{con_private}
\item $V_2\left(f\right)$ is a minimal dominating set on $G'$.\label{con_min_dom}
\end{enumerate}
\end{theorem}

Let $G=(V,E)$ be graph. Define the sets of functions $\mRdfSet{G}=\{f:V\to\{0,1,2\}\mid f$ is a minimal Roman dominating function$\,\}$ and $\mpRdfSet{G}=\{f:V\to\{0,1,2\}\mid f$ is a minimal perfect Roman dominating function$\,\}$.

\begin{theorem}\label{thm:bij_prdf_rdf}
Let $G=(V,E)$ be graph. 
There is a bijection~$B\colon \mRdfSet{G} \to \mRdfSet{G}$. Furthermore,  $B(f)$ and $B^{-1}(g)$ can be computed in polynomial time (with respect to $G$) for each $f\in \mRdfSet{G}$ and $g\in \mpRdfSet{G}$.  
\end{theorem}

\begin{pf}
Let $f\in \mRdfSet{G}$ and $g\in \mpRdfSet{G}$.
Define $B(f)$ by the three sets
\begin{equation*}
    \begin{split}
        V_0(B(f))=&\, \{v\in V_0(f) \mid \vert N_G(v) \cap V_2(f)\vert = 1 \},\\
        V_1(B(f))=&\, \{v\in V_0(f) \mid \vert N_G(v) \cap V_2(f)\vert \geq 2 \} \cup V_1(f),\\
        V_2(B(f))=&\, V_2(f).
    \end{split}
\end{equation*}
By definition of $B$, $\vert N_G(v) \cap V_2(f) \vert = 1$
holds for all $v\in V_0(B(f))$. Since each $v\in V$ with $f(v)=2$ needs a private neighbor except itself by \autoref{t_property_min_rdf}, $N_G(v) \cap V_0(B(f))\neq \emptyset$ holds. As any $v\in V_1(f)$ has no neighbor in $V_2(f)$ and any $v\in V_1(B(f))\setminus V_1(f)$ has at least two neighbors in $V_2(f)$, all conditions of \autoref{lem:conditionsPRDF} are met. Therefore, $B(f)\in \mpRdfSet{G}$.  

Define $B^{-1}(g)$ by the three sets
\begin{equation*}
    \begin{split}
        V_0(B^{-1}(g))=\,& V_0(g) \cup \{v\in V_1(g) \mid \vert N_G(v) \cap V_2(g)\vert \geq 2 \} ,\\
        V_1(B^{-1}(g))=\,& \{ v\in V_1(g) \mid \vert N_G(v) \cap V_2(g)\vert = 0 \},\\
        V_2(B^{-1}(g))=\,& V_2(f).
    \end{split}
\end{equation*} 
By definition of $B^{-1}$, $N_G(v)\cap V_2(B^{-1}(g)) = \emptyset$ for each $v\in V_1(B^{-1}(g))$. By \autoref{lem:conditionsPRDF}, each $v\in V_0(g)$ has exactly one neighbor in $V_2(g)$. Therefore, each vertex in $V_0(B^{-1}(g))$ has a neighbor in $V_2(B^{-1}(g))$ and $V_2(B^{-1}(g))$ is a dominating set on $G[V_0(B^{-1}(g)) \cup V_2(B^{-1}(g))]$. \autoref{lem:conditionsPRDF} implies that each $v\in V_2(g)$ has a $u\in N_G(v)\cap V_0(g)$ with $\{v\} = N_G(u) \cap V_2(g)$. Hence, $u$ is a private neighbor of~$v$ and $B^{-1}(g)\in \mRdfSet{G}$.

This leaves to show that $B^{-1}(B(f))=f$ and $B(B^{-1}(g)) = g$. Trivially,  $V_2(B^{-1}(B(f))) = V_2(f)$ and $V_2(B(B^{-1}(g))) = V_2(g)$ hold. Let $v\in V_1(f)$. By \autoref{t_property_min_rdf}, we know $ N_G(v)\cap V_2(f)=\emptyset$. This implies $v\in V_1(B(f))\cap V_1(B^{-1}(B(f)))$. Let $v\in V_0(f)$ with $\vert N_G(v)\cap V_2(f)\vert=1 $. Thus, $v\in V_0(B(f))\cap V_0(B^{-1}(B(f)))$. Let $v\in V_0(f)$ with $\vert N_G(v)\cap V_2(f)\vert \geq 2 $. Therefore, $v\in V_1(B(f))$ holds. The definition of $B^{-1}$ implies $v\in V_0(B^{-1}(B(f)))$. Hence, $B^{-1}(B(f))=f$. Assume $v\in V_0(g)$. \autoref{lem:conditionsPRDF} implies $\vert N_G(v)\cap V_2(f)\vert =1$. Therefore, $v\in V_0(B(B^{-1}(f)))\cap V_0(B^{-1}(f))$. For $v\in V_1(g)$ with $N_G(v)\cap V_2(g) = \emptyset$ the construction of the functions $B,B^{-1}$ implies $v\in V_1(B(B^{-1}(f)))\cap V_1(B^{-1}(f))$. Let $v\in V_1(g)$ with $\vert N_G(v)\cap V_2(g) \vert \geq 2$. This leads to $v\in V_0(B^{-1}(g))$ and $v\in V_1(B(B^{-1}(g)))$.  Hence, $B(B^{-1}(g))=g$. Therefore, the theorem holds.    
\end{pf}

\noindent
This bijection is a so-called \emph{parsimonious reduction}. This is a class of reductions designed for enumeration problems. For more information, we refer to \cite{Str2019}. Even without this paper, \autoref{thm:bij_prdf_rdf} implies some further results thanks to~\cite{AbuFerMan2022b}.

\begin{corollary}
There are graphs of order $n$ that have at least ${\sqrt[5]{16}\,}^n\in\Omega(\RomanLowerbound^n)$ many minimal perfect rdf.
\end{corollary}

\begin{theorem}\label{cor:minimal-prdf-enumeration}
There is a polynomial-space algorithm that enumerates all minimal perfect rdf of a given graph of order $n$ with polynomial delay and in time $\mathcal{O}^*(\RomanUpperbound^n)$.
\end{theorem}

According to~\cite{AbuFerMan2023}, we get some further results for special graph classes:
Let $G=(V,E)$, with $n \coloneqq \vert V \vert$ being the order of~$G$. If $G$ is a forest or interval graph, then there is a \prdf enumeration algorithm that runs in time $\mathcal{O}^*(\sqrt{3}^n)$ with polynomial delay. For both graph classes, a graph with many isolated edges is example of a graph with $\sqrt{3}^n$ many minimal \prdf. This is also the worst-case example for chordal graphs which is known so far. We can enumerate  all \prdfs of these graphs in time $\mathcal{O}(1.8940^n)$. If $G$ is a split or cobipartite graph, then we can enumerate all \prdfs of $G$ in $\mathcal{O}^*(\sqrt[3]{3}^n)$ with polynomial delay. Further, both classes include graphs of order~$n$ with  $\Omega(\sqrt[3]{3}^n)$ many \prdfs.
There is also a polynomial-time recursive algorithm to count all \prdfs of paths.

\begin{remark}\label{rem:minimum_rdf_vs_prdf}
    \begin{figure}[bt]
    \centering
	\begin{tikzpicture}[transform shape]
		      \tikzset{every node/.style={ fill = white,circle,minimum size=0.3cm}}
			\node[draw] (v1) at (-2,1) {$v_1$};
			\node[draw] (v2) at (-2,-1) {$v_2$};
			\node[draw] (v3) at (-1,0) {$v_3$};
			\node[draw] (v4) at (0,1) {$v_4$};
			\node[draw] (v5) at (0,-1) {$v_5$};
			\node[draw] (v6) at (1,0) {$v_6$};
			\node[draw] (v7) at (2,1) {$v_7$};
			\node[draw] (v8) at (2,-1) {$v_8$};
   			
            \path (v1) edge[-] (v3);
			\path (v2) edge[-] (v3);		
            \path (v5) edge[-] (v3);
			\path (v4) edge[-] (v3);		
            \path (v4) edge[-] (v6);
			\path (v5) edge[-] (v6);
			\path (v7) edge[-] (v6);
			\path (v8) edge[-] (v6);
        \end{tikzpicture}

        \label{fig:B_counterexample_minimum}
    \caption{Counter-example of \autoref{rem:minimum_rdf_vs_prdf}}
\end{figure}
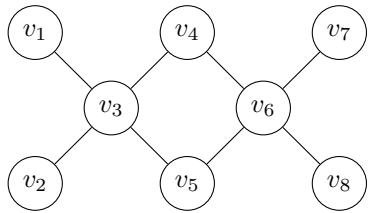
It should be mentioned that the function $B$ is inheriting the minimality but it does not mean that if $f$ is a minim\underline{um} Roman dominating function that $B(f)$ is a minim\underline{um} \prdf. For this we consider the graph $G=(V,E)$ with $V \coloneqq \{v_1, \ldots, v_8\}$ and 
$$E \coloneqq \{ \{v_1,v_3\},\{v_2,v_3\},\{v_4,v_3\},\{v_5,v_3\},\{v_4,v_6\}, \{v_5,v_6\}, \{v_7,v_6\}, \{v_8,v_6\}\}.$$ 
Since $v_1, v_2$ and $v_7,v_8$ are pairs of false twins and $v_1, v_7$ have no common neighbors, for each \rdf $f$ on $G$, $\weightfunction{f} \geq 4$. Therefore, $f\in \{0,1,2\}^V$ with $V_0(f)=\{v_1, v_2, v_4, v_5, v_7,v_8\}$, $V_1(f)=\emptyset $ and $V_2(f)=\{v_3,v_6\}$ is a minimum \rdf. $B(f)$ is given by $V_0(B(f))=\{v_1,v_2,v_7,v_8\}$, $V_1(B(f))=\{v_4,v_5\}$ and $V_2(f)=\{v_3,v_6\}$. The weight is $\weightfunction{B(f)}=6$. The minimal \prdf $g\in \{0,1,2\}$ with $V_0(g) = \{v_1, v_2, v_4, v_5\}$, $V_1(g) = \{v_6, v_7, v_8\}$ and $V_2(g) = \{v_3\}$ fulfills $\weightfunction{g} = 5$. Hence, $B(f)$ is no minimum \prdf.

Let $h\in \{0,1,2\}^V$ be minimal \prdf. For $h(v_3)=h(v_6)=2$, $h=B(f)$, as $v_1,v_2,v_7,v_8 \in N(V_2(h))$ are pendant. Assume $h(v_3) = 2 \neq  h(v_6)$. If $h(v_4)= 2$ (respectively $h(v_5)= 2$), then $h(v_7)=h(v_8)=1$ and $\weightfunction{h} \geq 6$. For $h(v_7)=2$ (respectively $h(v_8)= 2$), $h(v_8)=1$ (respectively $h(v_7)= 1$) and $\weightfunction{g}=5\geq \weightfunction{h}$. As $v_1,v_2\notin V_2(h)$ for $h(v_3)=2$, the remaining possibility is $h=g$. Thus, $\weightfunction{g}=5\leq \weightfunction{h}$ holds for each minimal \prdf with $h(v_3) = 2 \neq h(v_6)$. Analogously, for \prdfs $h$ with $h(v_6)\neq 2 = h(v_3)$, $\weightfunction{g}=5\leq \weightfunction{h}$. Let $h(v_3)\neq 2$ and $h(v_6)\neq 2$. Since $N(v_i) \subseteq \{v_3,v_6\}$ for $i\in \{1,2,4,5,7,8\}$, $v_1,v_2,v_4,v_5,v_7,v_8 \notin V_0(h)$. Hence, $\weightfunction{h}\geq 6$ and $g$ is a minimum \prdf. Furthermore, $B^{-1}(g) = g$ is not a minimum \rdf.     
\end{remark}

\section{Conclusion}

We presented polynomial-time algorithms for \URRD\ on cobipartite and split graphs and one for \PRD\ on cobipartite graphs. On split graphs, \PRD\ is \NP-complete but we provided an \FPT-algorithm, parameterized by solution size. Then we gave an $\mathcal{O}^*\left(\sqrt[3]{3}^n\right)$
enumeration algorithm for \urrdfs on graphs of order $n$ without isolated vertices. This is an optimal algorithm as we also found a family of graphs without isolated vertices of order $n$ and $\sqrt[3]{3}^n$ many \urrdfs. Although the extension version of \PRD\ is \NP-complete, proven to be even \W{1}-complete if parameterized by $\weightfunction{f}$ and \W{2}-complete if parameterized by $\vert V_0(f) \vert$, we showed that all \prdfs of a graph of order $n$ can be enumerated in $\mathcal{O}^*(1.9332^n)$ with polynomial delay. This is interesting, as most often polyomial delay is linked to a polynomial-time decision algorithm for the extension version of the corresponding property. The main technique is to devise bijections to objects that can be enumerated with polynomial delay, as (in our case) minimal \rdfs. This technique can be also applied to enumerate all minimal unique response strong Roman dominating functions, as introduced in \cite{MojHMP2021}.

It could also be interesting to consider enumeration algorithms for other variations of Roman domination functions such as double/connected/total Roman dominating functions. In particular, it would be interesting to study under which circumstances polynomial-delay enumeration is possible.
Another interesting research direction is to look into other graph classes which we did not consider in this paper, now focusing on an input-sensitive analysis. For example, we do not know of any enumeration algorithm for minimal \rdfs/\prdfs on bipartite graphs which is better than the general one, although this class looks similar to the ones of bipartite and split graphs where we could achieve considerable improvements over the general case, see~\cite{AbuFerMan2023}.


\bibliographystyle{splncs04}
\bibliography{ab,hen}
\end{document}